\documentclass[review]{elsarticle}

\usepackage{lineno,hyperref} \usepackage{mathtools, bm} \usepackage{amssymb, bm}
\usepackage{float} \modulolinenumbers[5]

\journal{Journal of \LaTeX\ Templates}

\usepackage{xcolor} 
\usepackage{xcolor} 

\begin{document}

\begin{frontmatter}

\title{A multi-dimensional, robust, and cell-centered finite-volume scheme for
  the ideal MHD equations}

\author[affiliationMdls]{Pascal Tremblin\corref{mycorrespondingauthor}}
\cortext[mycorrespondingauthor]{Corresponding author}
\ead{pascal.tremblin@cea.fr} \author[affiliationMdls]{Rémi Bourgeois}
\author[affiliationMdls]{Solène Bulteau} \author[affiliationSTMF]{Samuel Kokh}
\author[affiliationMdls]{Thomas Padioleau}
\author[affiliationDap]{Maxime Delorme}
\author[affiliationDap]{Antoine Strugarek}
\author[affiliationDap2]{Matthias Gonz\'alez}
\author[affiliationDap]{Allan Sacha Brun}

\address[affiliationMdls]{Universit\'e Paris-Saclay, UVSQ, CNRS, CEA, Maison de
  la Simulation, 91191, Gif-sur-Yvette, France}
\address[affiliationSTMF]{Universi\'e Paris-Saclay, CEA,  Service de G\'enie
  Logiciel pour la Simulation, 91191, Gif-sur-Yvette, France}
\address[affiliationDap]{Universit\'e Paris-Saclay, Universit\'e Paris Cit\'e, CEA,
  CNRS, AIM, 91191, Gif-sur-Yvette, France}
\address[affiliationDap2]{Universite\'e Paris Cite\'e, Universite\'e Paris-Saclay,
  CEA, CNRS, AIM, F-91191, Gif-sur-Yvette, France}

\begin{abstract}
We present a new multi-dimensional, robust, and cell-centered finite-volume
scheme for the ideal MHD equations.  This scheme relies on relaxation and
splitting techniques and can be easily used at high order. A fully conservative
version is not entropy satisfying but is observed experimentally to be more
robust than standard constrained transport schemes at low plasma beta. At very
low plasma beta and high Alfvén number, we have designed an entropy-satisfying
version that is not conservative for the magnetic field but preserves admissible
states and we switch locally a-priori between the two versions depending on the
regime of plasma beta and Alfvén number. This strategy is robust in a wide range
of standard MHD test cases, all performed at second order with a classic
MUSCL-Hancock scheme.
\end{abstract}


\end{frontmatter}

\renewcommand{\vector}[1]{\ensuremath{\mathbf{{#1}}}}

  \section{Introduction}
Developing a robust multi-dimensional numerical scheme for the ideal MHD
equations remains a challenge that is of great importance for astrophysics and
plasma physics applications. A MHD flow is characterized by an exact
zero-divergence magnetic field, and by using terms that are proportional to the
divergence of the magnetic field, the MHD equations can be put in a fully
conservative form, with density, momentum, energy and magnetic field
conservation. However this form introduces a source term proportional to the
divergence of the magnetic field on the entropy evolution equation, leading to
an unstable scheme for multi-dimensional test cases, because of discretization
errors on this source term.

A solution to this problem is to remove the divergence errors so that the source term in
the entropy evolution equation is as small as possible. Such a solution encompass
the divergence-cleaning method (see \cite{brackbill:1980,ryu:1998,dai:1998,dedner:2001}) and
the constrained transport method (see
\cite{evans:1988,balsara:1999,TOTH2000605,fromang:2006}). These methods greatly
improve the stability of MHD numerical schemes and has been used in numerous
applications in astrophysics and plasma physics. However, they are not entropy
satisfying and may fail with negative energies especially in the low plasma beta
regime. This problem is mitigated by using a threshold value for the internal
energy, effectively breaking the energy conservation of the numerical scheme
(ref divB cleaning with threshold).
Another solution is to design an entropy satisfying numerical
scheme for any value of the divergence of the magnetic field.  This solution has
been explored using relaxation methods in \cite{gallice:2003,bouchut:2007,bouchut:2010}.
Originally, it has been shown that a multi-dimensional solver with the
introduction of non-conservative Powell source terms in the momentum, induction
and energy evolution equations allows to obtain a symmetric form of the MHD equations
\cite{godunov:1972,busto:2023}, but \cite{bouchut:2010} has demonstrated that it
is also possible to obtain a symmetric form with a source term only on the induction
equation, therefore preserving energy and momentum conservation with an entropy
satisfying numerical scheme.

In recent years, significant advancements have been made in splitting strategies
for designing numerical solvers for Euler equations. In the works by \cite{chalons:2014,
  padioleau:2019,bourgeois:2024}, the
approximation algorithm is divided into two steps: an acoustic step and a
transport step. For one-dimensional cases, these methods resemble the explicit
Lagrange-Projection approach
\citep{hirt:1974,godlewski1996,despres:2010}. However, this new splitting
technique avoids the use of a moving
Lagrangian mesh and is readily adaptable to multi-dimensional problems.
On the other hand, significant progresses have also been made on approximate
Riemann solvers based on relaxation strategies
\citep{jin:1995,suliciu:1998,coquel:2001,bouchut:2004,chalons:2008,coquel:2010}.

In this paper, we build on the proposition of a relaxation approximation for the
MHD system \cite{bouchut:2007,bouchut:2010} by taking advantage
of splitting techniques introduced in \cite{chalons:2014} to design a
fully-conservative multi-dimensional MHD solver in regions of high plasma beta /
low Alfvén number, and an entropy satisfying version with an entropy correction
in regions of low plasma beta / high Alfvén number.  The resulting solver,
therefore, allies a
robust entropy-satisfying and a fully-conservative scheme depending on the regime
of the flow. In Sect. 2, we
introduce the different systems of equations for MHD (conservative and
non-conservative) and the corresponding entropy evolution equation. In Sect. 3,
we present the splitting of the equations in a magneto-acoustic and transport
subsystems. Sect. 4 and 5 introduce the numerical methods used to solve the
evolution of these subsystems and Sect. 6 provides the global fully-conservative
scheme for the MHD system of equations. Sect. 7 is devoted to the entropy
analysis of the numerical method, showing that the fully-conservative solver is
not entropy-satisfying and introduces an entropy correction on the induction
equation in order to obtain
an entropy-satisfying method at the price of loosing the magnetic field
conservation. In Sect. 8, we provide numerical tests in 1D and 2D at second
order by leveraging the advantages of the fully-conservative and
entropy-satisfying solvers depending on the regime of the flow. We provide our
conclusions and a discussion in Sect. 9.

\section{MHD equations}

The ideal MHD equations are given by the evolution equations of the fluid
density $\rho$, momentum $\rho \vector{u}$, energy $\rho (e+\vector{u}^2/2)$,
and the Faraday's law of induction describing the evolution of the magnetic
field $\vector{B}$

\begin{eqnarray}\label{eq:mhd_noncons}
  \partial_t \rho + \vector{\nabla}\cdot(\rho\vector{u}) = 0, \cr
  \partial_t(\rho \vector{u}) +\vector{\nabla}\cdot(\rho
  \vector{u}\otimes\vector{u} ) = -\vector{\nabla}p + \vector{j} \times
  \vector{B},\cr \partial_t(\rho (e+\vector{u}^2/2)) +
  \vector{\nabla}\cdot(\rho(e+\vector{u}^2/2)\vector{u}) = -
  \vector{\nabla}\cdot(p\vector{u})+(\vector{j}\times\vector{B})\cdot\vector{u},\cr
  \partial_t \vector{B} + \vector{\nabla}\times \vector{E} = 0.
\end{eqnarray}
The term $\vector{j}\times\vector{B}$ is the Lorentz force.  This system of
equations is closed with the ideal Ohm's law $\vector{E} = -
\vector{u}\times\vector{B}$, the low frequency Maxwell equation $\vector{j} =
\vector{\nabla}\times\vector{B}$ assuming a system of units in which the vacuum
permeability is one, and an equation of state connecting the pressure $p$ to the
density $\rho$ (or specific volume $\tau=1/\rho$) and internal energy $e$. The
equation of state also defines the specific physical entropy $s(\tau,e)$
assuming that $-s$ is a convex function of $(\tau,e)$, and satisfies
\begin{equation}\label{eq:thermo}
de+pd\tau = Tds.
\end{equation}
This equivalently means that the internal energy is convex with respect
  to specific volume and entropy, hence the sound speed $c_s$ defined by
\begin{equation}
c_s^2 = \left(\frac{\partial p }{\partial \rho}\right)_s
\end{equation}
is positive and ensures the hyperbolicity of the system. Assuming smooth solutions
of (\ref{eq:mhd_noncons}), one can show that they satisfy the following equation
of conservation for the entropy
\begin{equation}
  \partial_t(\rho s) + \vector{\nabla}\cdot(\rho s\vector{u}) = 0.
\end{equation}
For the non-conservative form of the MHD equations, this holds for any value of
the divergence of the magnetic field $\vector{\nabla}\cdot \vector B$.  Assuming
that the divergence of the magnetic field is zero at an initial time
$\vector{\nabla}\cdot\vector{B} = 0$, it remains zero at all time following the
divergence of the induction equation,
\begin{equation}
\partial_t(\vector{\nabla}\cdot\vector{B}) = 0.
\end{equation}
The free divergence constraint is therefore a consequence of the induction
equation and not a dynamical constraint.

Equivalently, by adding terms proportional to $\vector{\nabla}\cdot\vector{B}$ in
the momentum and energy equations (see
  \ref{sect:nonconstocons}), one can obtain a conservative form
for the MHD equations
\begin{eqnarray}\label{eq:mhd_cons}
  \partial_t \rho + \vector{\nabla}\cdot(\rho\vector{u}) = 0, \cr
  \partial_t(\rho \vector{u}) +\vector{\nabla}\cdot(\rho
  \vector{u}\otimes\vector{u} +\vector{\sigma} -\vector{B}\otimes\vector{B} ) =
  0,\cr \partial_t(\rho E)+\vector{\nabla}\cdot(\rho
  E\vector{u}+\sigma\cdot\vector{u}- (\vector{B}\cdot\vector{u})\vector{B}) =
  0,\cr \partial_t \vector{B} + \vector{\nabla}\cdot(\vector{u}\otimes\vector{B}
  -\vector{B}\otimes\vector{u})= 0.
\end{eqnarray}
with $\vector{\sigma} = (p+\vector{B}^2/2)\vector{I}$ and $E =
e+\vector{u}^2/2+\vector{B}^2/(2\rho)$.  Assuming smooth solutions of
(\ref{eq:mhd_cons}), one can show that they satisfy the following equation for the
evolution of the entropy by subtracting the evolution of the kinetic and
  magnetic energy from the evolution of the total energy
\begin{equation}\label{eq:entropy_cons}
  \partial_t(\rho s) + \vector{\nabla}\cdot(\rho s\vector{u}) =
  -\frac{\vector{u}\cdot\vector{B}}{T} \vector{\nabla}\cdot \vector{B},
\end{equation}
which is compatible with entropy conservation only when $\vector{\nabla}\cdot
\vector{B}=0$ in constrast to the non-conservative form presented above \cite{despre:2011}. This
shows that the entropy balance is closely related to the free divergence
constraint for the conservative MHD equations.

In the case of discontinuities such as shocks and in order to ensure
dissipation, the second law of thermodynamics must be enforced and implies the
entropy inequality
\begin{equation}\label{eq:entropy_scheme}
 \partial_t(\rho s) + \vector{\nabla}\cdot(\rho s\vector{u}) \ge 0,
\end{equation}
 After discretization, truncation errors on the
$\vector{\nabla}\cdot\vector{B}$ source term in Eq. (\ref{eq:entropy_cons})therefore
leads to some issues in order to obtain an entropy satisfying numerical scheme
ensuring a discrete version of Eq. (\ref{eq:entropy_scheme}).

In the next sections (3, 4, 5 and 6), we introduce a new fully-conservative
solver relying on a splitting between a magneto-acoustic and a transport
subsystem. This solver is entropy satisfying and is not compatible with
Eq. (\ref{eq:entropy_scheme}). We then introduce in Sect. 7 an entropy correction
following \cite{bouchut:2010} ensuring that the modified scheme is compatible with
Eq. (\ref{eq:entropy_scheme}) while breaking the magnetic field conservation but
maintaining the momentum and energy conservation.

\section{Magneto-acoustic/transport splitting}

Similarly to \cite{chalons:2014}, we propose the following splitting of the
conservative MHD equations into a magneto-acoustic sub-system
\begin{eqnarray}\label{eq:mag-ac}
  \partial_t \rho + \rho\vector{\nabla}\cdot\vector{u} = 0, \cr \partial_t(\rho
  \vector{u}) + \rho\vector{u}\vector{\nabla}\cdot\vector{u} +
  \vector{\nabla}\cdot(\vector{\sigma} -\vector{B}\otimes\vector{B} ) = 0,\cr
  \partial_t(\rho E)+\rho E\vector{\nabla}\cdot\vector{u} +
  \vector{\nabla}\cdot(\vector{\sigma}\cdot\vector{u}-
  (\vector{B}\cdot\vector{u})\vector{B}) = 0,\cr \partial_t \vector{B} +
  \vector{B}\vector{\nabla}\cdot\vector{u}-\vector{\nabla}\cdot(
  \vector{B}\otimes\vector{u})= 0,
\end{eqnarray}
and a transport sub-system
\begin{eqnarray}\label{eq:trans}
  \partial_t \rho + \vector{u}\cdot\vector{\nabla}\rho = 0, \cr \partial_t(\rho
  \vector{u}) + \vector{u}\cdot\vector{\nabla}(\rho\vector{u}) = 0,\cr
  \partial_t(\rho E)+\vector{u}\cdot\vector{\nabla}(\rho E)=0,\cr \partial_t
  \vector{B} + \vector{u}\cdot\vector{\nabla}(\vector{B})= 0.
\end{eqnarray}
We emphasize that all the components of the magnetic field are transported at
velocity $\vector{u}$ in the transport sub-system.  We then propose to
approximate the solution of Eq. (\ref{eq:mhd_cons}) by approximating the solutions of
the two sub-systems (\ref{eq:mag-ac}) and (\ref{eq:trans}), i.e.  for a discrete
state $\vector{U}^n_i = (\rho,\rho\vector{u},\rho E, \vector{B})^n_i$ in a cell
$\Omega_i$ at time $t^n$, the update to $\vector{U}^{n+1}_i$ is first an update
from $\vector{U}^{n}_i$ to $\vector{U}^{n+1-}_i$ by approximating the solution
of (\ref{eq:mag-ac}), then an update from $\vector{U}^{n+1-}_i$ to
$\vector{U}^{n+1}_i$ by approximating the solution of (\ref{eq:trans}).  We
present in Sect.~\ref{sect:mag-ac} and in Sect.~\ref{sect:trans} the
discretization and the entropy analysis for each sub-system respectively.

\section{Relaxation approximation of the magneto-acoustic sub-system}
\label{sect:mag-ac}

The relaxation approximation of the magneto-acoustic sub-system and the
  associated entropy analysis in Sect. \ref{sect:entropy} heavily
  relies on earlier works by \cite{bouchut:2007, bouchut:2010}. We highlight two
main differences in our approach: we keep in the analysis gradients of the
magnetic field perpendicular to the interface that appears in the multi-dimensional
case and we propose a different choice of relaxation parameters in the 5-wave
solver to ensure the strict hyperbolicity of the relaxed system.

The multi-dimensional scheme will be obtained by taking advantage of the
rotational invariance of the magneto-acoustic sub-system, following the lines of
\cite{godlewski1996}. We, therefore, rewrite sub-system (\ref{eq:mag-ac}) in 1D,
and simplify it by using the density evolution equation
\begin{eqnarray}\label{eq:mag-ac_1D}
  \rho\partial_t \tau - \partial_x u = 0, \cr \rho\partial_t
  \vector{u}+\partial_x(\sigma\vector{e}_x- B_x\vector{B} ) = 0,\cr \rho\partial_t
  E+\partial_x(\sigma u_x- (\vector{B}\cdot\vector{u}) B_x) = 0,\cr
  \rho\partial_t (\tau\vector{B}) -\partial_x(B_x \vector{u})= 0,
\end{eqnarray}
with $\vector{e}_x$, the unit vector normal to the interface, $B_x$,
  $B_y$, and $B_z$ the components of the magnetic field and $u_x$,
  $u_y$, and $u_z$ the components of the velocity field.  The eigenvalues of
this sub-system are given by
\begin{equation}
-u,0,\pm c_{ms}, \pm c_{ma}, \pm c_{mf}
\end{equation}
with $c_{ma}$, the magnetic Alfvén speed, $c_{ms}$, the slow magnetosonic speed,
$c_{mf}$, the fast magnetosonic speed defined by
\begin{eqnarray}
  c_{ma} &=& \frac{|B_x|}{\sqrt{\rho}}, \cr c_{ms}^2 &=&
  \frac{1}{2}\left(c_s^2+\frac{\vector{B}^2}{\rho} -
  \sqrt{\left(c_s^2+\frac{\vector{B}^2}{\rho}\right)^2 - 4 c_s^2
    c_{ma}^2}\right), \cr c_{mf}^2 &=&
  \frac{1}{2}\left(c_s^2+\frac{\vector{B}^2}{\rho} +
  \sqrt{\left(c_s^2+\frac{\vector{B}^2}{\rho}\right)^2 - 4 c_s^2
    c_{ma}^2}\right).
\end{eqnarray}

We then introduce a relaxation procedure \cite{bouchut:2007, chalons:2014} with
the relaxation pressures $\pi_\vector{u}$
playing the role of the fluxes in the impulsion equation and the relaxation
variable $r$ playing the role of the density in front of the time derivatives
\begin{eqnarray}\label{eq:relax1}
  r\partial_t \tau - \partial_x u = 0, \cr r\partial_t \vector{u}+\partial_x
  \pi_\vector{u} = 0,\cr r\partial_t E+\partial_x(\pi_\vector{u}\cdot\vector{u})
  = 0,\cr r\partial_t (\tau\vector{B}) -\partial_x(B_x \vector{u})= 0,
\end{eqnarray}
with the following equations for the relaxation variables
\begin{eqnarray}\label{eq:relax2}
  \partial_t r = \frac{\rho-r}{\epsilon}, \cr r\partial_t\pi_u+
  (c_b^2+b_y^2+b_z^2)\partial_x u -c_a b_y \partial_x v -c_a b_z \partial_x w
  +d_x \partial_x B_x= \frac{\sigma-B_x^2 - \pi_u}{\epsilon},\cr
  r\partial_t\pi_v -c_a b_y \partial_x u + c_a^2\partial_x v +d_y \partial_x
  B_x= \frac{-B_xB_y - \pi_v}{\epsilon},\cr r\partial_t\pi_w -c_a b_z \partial_x
  u + c_a^2\partial_x w +d_z \partial_x B_x= \frac{-B_xB_z - \pi_w}{\epsilon}.
\end{eqnarray}
The parameters $c_a$, $c_b$, $b_y$, $b_z$ play the role of approximations of
$\sqrt{\rho}|B_x|$, $\rho c_s$, $\mathrm{sign}(B_x)\sqrt{\rho}B_y$,
$\mathrm{sign}(B_x)\sqrt{\rho}B_z$, respectively, as in \cite{bouchut:2007}. The
extra parameters $d_x$, $d_y$, $d_z$ are linked to the possibility of a
non-constant $B_x$ in the magneto-acoustic sub-system and play the role of
approximations of $2B_xu/\tau+ \vector{u}\cdot\vector{B}(\partial_e p-1/\tau)$,
$(B_x v+ B_y u)/\tau$, and $(B_x w + B_z u)/\tau$, respectively. If these extra
parameters are fixed to zero, the relaxation equations for $\pi_\vector{u}$ is
the Lagrangian form of the relaxation equations used in \cite{bouchut:2007}.
By replacing all these parameters exactly by the quantities they
  approximate, Eq.~(\ref{eq:relax2}) reduces to the evolution equation of
  $\sigma-B_x^2$, $-B_x B_y$, and $-B_x B_z$ in the limit  $\epsilon \rightarrow
  \infty$.
In
order to obtain the same Riemann invariants as \cite{bouchut:2007}, we fix
$d_x$, $d_y$, and $d_z$ to zero and the other constants are evolved with
\begin{equation}
\partial_t c_a = \partial_t c_b = \partial_t b_y = \partial_t b_z =0.
\end{equation}
In the limit $\epsilon \rightarrow 0$, the relaxation equations in
Eq.~\ref{eq:relax2} ensures that $r \rightarrow \rho$,  $\pi_u \rightarrow
\sigma-B_x^2$, $\pi_v \rightarrow -B_x B_y$, and $\pi_w \rightarrow -B_x
B_z$. In this limit, Eq.~(\ref{eq:relax1}) is then equivalent to
Eq.~(\ref{eq:mag-ac_1D}). A classical approach to achieve the limit $\epsilon
\rightarrow 0$ numerically is to first
enforce the equilibrium relations $r=\rho$ and $\pi_{\vector{u}} =
\sigma\vector{e}_x-B_{x} \vector{B}$ at time $t^n$ and then solve (\ref{eq:relax1})
and (\ref{eq:relax2}) without the relaxation source terms. Using $L\equiv r/\rho$,
the full system without the relaxation source term is

\begin{eqnarray}\label{eq:full_relax}
  \partial_t L - \partial_x u = 0, \cr \partial_t(\rho L \vector{u})+\partial_x
  \pi_\vector{u} = 0,\cr \partial_t(\rho L
  E)+\partial_x(\pi_\vector{u}\cdot\vector{u}) = 0,\cr \partial_t (L\vector{B})
  -\partial_x(B_x \vector{u})= 0,\cr \partial_t(\rho L) = 0, \cr \partial_t(\rho
  L \pi_u)+ (c_b^2+b_y^2+b_z^2)\partial_x u -c_a b_y \partial_x v -c_a b_z
  \partial_x w = 0,\cr \partial_t(\rho L \pi_v) -c_a b_y \partial_x u +
  c_a^2\partial_x v = 0,\cr \partial_t(\rho L \pi_w) -c_a b_z \partial_x u +
  c_a^2\partial_x w = 0.
\end{eqnarray}
After some tedious algebra, one can compute the eigenvalues of this system of 16
equations (including $\partial_t c_a = \partial_t c_b = \partial_t b_y =
\partial_t b_z =0$),
\begin{equation}
-u/L,0,\pm c_{rs}/(\rho L), \pm c_{ra}/(\rho L), \pm c_{rf}/(\rho L)
\end{equation}
with
\begin{eqnarray}
  c_{ra} &=& c_a, \cr c_{rs}^2 &=& \frac{1}{2}\left(c_b^2+c_a^2+b_y^2+b_z^2 -
  \sqrt{\left(c_b^2+c_a^2+b_y^2+b_z^2\right)^2 - 4 c_a^2 c_{b}^2}\right), \cr
  c_{rf}^2 &=& \frac{1}{2}\left(c_b^2+c_a^2+b_y^2+b_z^2 +
  \sqrt{\left(c_b^2+c_a^2+b_y^2+b_z^2\right)^2 - 4 c_a^2 c_{b}^2}\right).
\end{eqnarray}
The central wave at zero velocity has multiplicity 9.  All the waves are
linearly degenerate.
Similarly to \cite{bouchut:2007}, $c_{rs}\le c_a \le c_{rf}$, $c_{rs}\le c_b \le
c_{rf}$ and the eigenvalues of (\ref{eq:full_relax}) match the eigenvalues of
(\ref{eq:mag-ac_1D}) for $c_a=\sqrt{\rho}|B_x|$, $c_b=\rho c_s$,
$b_y=\mathrm{sign}(B_x)\sqrt{\rho}B_y$, $b_z=\mathrm{sign}(B_x)\sqrt{\rho}B_z$.
Similarly to \cite{bouchut:2007}, a Chapman-Enskog analysis can be performed on
the relaxation equations which leads to the following stability conditions
\begin{eqnarray}
  \frac{1}{\rho}-\frac{B_x^2}{c_a^2}&\ge& 0,\cr c_b^2-\rho^2 c_s^2 &\ge& 0,\cr
  (c_b^2-\rho^2 c_s^2)\left( \frac{1}{\rho}-\frac{B_x^2}{c_a^2}\right)&\ge&
  \left(B_y-\frac{B_x b_y}{c_a}\right)^2+\left(B_z-\frac{B_x b_z}{c_a}\right)^2,
\end{eqnarray}
in order to ensure positive eigenvalues of the entropy diffusion matrix.

\paragraph{The 3+1 and 5+1 wave solver}
The solution of the Riemann problem associated to (\ref{eq:full_relax}) contain
7+1 waves in the general case, 7 waves that are identical to a Lagrangian
version of the 1D relaxation solver presented in \cite{bouchut:2007} to which we
add a wave at $-u/L$ associated to $B_x$.  Similarly to \cite{bouchut:2010} we can
design an approximate Riemann solver with 5+1 waves by choosing $b_y=b_z=0$, or
with 3+1 waves by choosing in addition $c_a=c_b=c$. The 5+1 wave solver is a
good compromise between accuracy ad computational cost and we will use this
approximation for now on.

We now look for strong Riemann invariants for the different waves by finding
quantities transported at the corresponding wave speed \cite{godlewski1996}.
$B_x$ is a strong Riemann invariant associated to the wave at $-u/L$. Note that
$B_x$ is not constant but advected at velocity $-u/L$.  $B_x$ has to be
understood as evaluated locally, upwind relative to the wave $-u/L$.  $c_a$ and
$c_b$ are strong Riemann invariants for the central wave with
\begin{equation}\label{eq:invariants}
\frac{1}{\rho}+\frac{\pi_u}{c_b^2}, \frac{B_y}{\rho}+\frac{B_x}{c_a^2}\pi_v,
\frac{B_z}{\rho}+\frac{B_x}{c_a^2}\pi_w,
e+\frac{\vector{B}^2}{2\rho}-\frac{\pi_u^2}{2c_b^2}-\frac{\pi_v^2+\pi_w^2}{2c_a^2}.
\end{equation}
Similarly to \cite{bouchut:2010}, there are six strong Riemann invariants for the
left and right waves $\pi_\vector{u}+c_\vector{u}\vector{u}$ and
$\pi_\vector{u}-c_\vector{u}\vector{u}$, respectively, in which we have defined
$c_\vector{u} = (c_b,c_a,c_a)$. Strong Riemann invariants for a given
  wave are weak Riemann invariants for the other waves. They are, therefore, weak
Riemann invariants for the central
wave, hence, $\vector{u}$ and $\pi_\vector{u}$ take the same value on the left
and right of this wave that we shall define as $\vector{u}^*$ and
$\pi^*_\vector{u}$ respectively. By using the weak Riemann invariants, we get
\begin{eqnarray}
\vector{u}^* &=& \frac{c_{\vector{u},l}\vector{u}_l +
  c_{\vector{u},r}\vector{u}_r + \pi_{\vector{u},l}-\pi_{\vector{u},r}}
       {c_{\vector{u},l}+c_{\vector{u},r}},\cr \pi_\vector{u}^* &=&
       \frac{c_{\vector{u},r}\pi_{\vector{u},l} +
         c_{\vector{u},l}\pi_{\vector{u},r} + c_{\vector{u},l}c_{\vector{u},r}
         (\vector{u}_l-\vector{u}_r)}{c_{\vector{u},l}+c_{\vector{u},r}}.
\end{eqnarray}
Then one has
\begin{equation}\label{eq:bx}
  B_x(x,t) = \left\{
  \begin{array}{ll}
    B_{x,l} & \mbox{if } x/t < -u/L \\ B_{x,r} & \mbox{if } x/t > -u/L, \\
  \end{array}
    \right.
\end{equation}
hence, at the interface, we define $B_x^{-u^*} = B_x(0,t)$ with
\begin{equation}
  B_x^{-u^*} = \left\{
  \begin{array}{ll}
    B_{x,l} & \mbox{if } u^* < 0 \\ B_{x,r} & \mbox{if } u^* > 0. \\
  \end{array}
    \right.
\end{equation}
The other intermediate states, e.g.  $\tau^*_{l,r}$ and $e^*_{l,r}$ can be
obtained by using (\ref{eq:invariants}), but are not needed for deriving
  the update of the numerical scheme.  The discrete numerical scheme for the
magneto-acoustic sub-system is then given by
\begin{eqnarray}\label{eq:num}
  L_i^{n+1-} &=& 1+\frac{\Delta t}{\Delta x}(u^*_{i+1/2}-u^*_{i-1/2}), \cr
  \rho_i^{n+1-} L_i^{n+1-} &=& \rho_i^{n}, \cr \rho_i^{n+1-}\vector{u}^{n+1-}_i
  L_i^{n+1-} &=& \rho_i^{n}\vector{u}^{n}_i - \frac{\Delta t}{\Delta
    x}(\pi^*_{\vector{u},i+1/2}-\pi^*_{\vector{u},i-1/2}),\cr
  \rho_i^{n+1-}E^{n+1-}_i L_i^{n+1-} &=& \rho_i^{n}E^{n}_i - \frac{\Delta
    t}{\Delta x}(\pi^*_{\vector{u},i+1/2}\cdot\vector{u}^*_{i+1/2} -
  \pi^*_{\vector{u},i-1/2}\cdot\vector{u}^*_{i-1/2}),\cr \vector{B}_i^{n+1-}
  L_i^{n+1-} &=& \vector{B}^n_i + \frac{\Delta t}{\Delta
    x}(B_{x,i+1/2}^{-u^*}\vector{u}^*_{i+1/2}-B_{x,i-1/2}^{-u^*}\vector{u}^*_{i-1/2}),
\end{eqnarray}
with the CFL condition for this scheme
\begin{equation}
\max_{i \in \mathbb{Z}}\left( \frac{c_{rf,i}}{\rho_i}\right) \Delta t \le
\frac{\Delta x}{2}.
\end{equation}

\section{Transport sub-system}\label{sect:trans}

The transport sub-system is a quasi-hyperbolic system that only involves the
transport of conservative variables with the velocity $u$.  We choose to
approximate the solution of the 1D version of (\ref{eq:trans}) thanks to a
standard upwind Finite-Volume approximation for $\vector{U} =
(\rho,\rho\vector{u},\rho E, \vector{B})$ by discretizing
\begin{equation}
\frac{\partial \vector{U}}{\partial t} + u \frac{\partial \vector{U}}{\partial
  x} =  \frac{\partial \vector{U}}{\partial t} + \frac{\partial
  (u\vector{U})}{\partial x} - \vector{U} \frac{\partial u}{\partial x} =0, 
\end{equation}
with
\begin{equation}
\vector{U}^{n+1}_i = \vector{U}^{n+1-}_i - \frac{\Delta t}{\Delta x}
(u^*_{i+1/2}\vector{U}_{i+1/2}-u^*_{i-1/2}\vector{U}_{i-1/2}) + \frac{\Delta
  t}{\Delta x} \vector{U}^{n+1-}_i (u^*_{i+1/2}-u^*_{i-1/2}),
\end{equation}
with two possible choices of discretization for the interface states $\vector{U}_{i-1/2}$ and
$\vector{U}_{i+1/2}$. The first choice
\begin{equation}
  \vector{U}_{i+1/2} = \left\{
  \begin{array}{ll}
    \vector{U}^{n+1-}_{i} & \mbox{if } u^*_{i+1/2} \ge 0,
    \\ \vector{U}^{n+1-}_{i+1} & \mbox{if } u^*_{i+1/2} \le 0, \\
  \end{array}
    \right.
\end{equation}
leads to a magneto-acoustic+transport scheme of stencil 2 similar to
\cite{chalons:2014}. The second choice
\begin{equation}
  \vector{U}_{i+1/2} = \left\{
  \begin{array}{ll}
    \vector{U}^{n}_{i} & \mbox{if } u^*_{i+1/2} \ge 0, \\ \vector{U}^{n}_{i+1} &
    \mbox{if } u^*_{i+1/2} \le 0, \\
  \end{array}
    \right.
\end{equation}
leads to a magneto-acoustic+transport scheme of stencil 1 similar to
\cite{bourgeois:2024}. We will refer to these choices of discretization
  as ``stencil 1'' and ``stencil 2'' in the rest of the paper.
In both cases and using the notation $u^\pm =
\frac{u\pm|u|}{2}$, the CFL condition of the transport sub-system is given by
\begin{equation}
\max_{i \in \mathbb{Z}}( (u^*_{i-1/2})^+-(u^*_{i+1/2})^-) \Delta t \le \Delta x.
\end{equation}
The transport can also be written in the form
\begin{equation}
\vector{U}^{n+1}_i = \vector{U}^{n+1-}_iL^{n+1-}_i - \frac{\Delta t}{\Delta x}
(u^*_{i+1/2}\vector{U}_{i+1/2}-u^*_{i-1/2}\vector{U}_{i-1/2}).
\end{equation}

\section{Magneto-acoustic+transport scheme}
The global scheme is given by
\begin{eqnarray}\label{eq:num}
  \rho_i^{n+1} = \rho_i^{n} &-& \frac{\Delta t}{\Delta x}
  (\rho_{i+1/2}u^*_{i+1/2}-\rho_{i-1/2}u^*_{i-1/2}), \cr
  (\rho\vector{u})^{n+1}_i = (\rho\vector{u})^{n}_i &-&\frac{\Delta t}{\Delta
    x}((\rho\vector{u})_{i+1/2}u^*_{i+1/2}+\pi^*_{\vector{u},i+1/2}\cr
  &&-(\rho\vector{u})_{i-1/2}u^*_{i-1/2}-\pi^*_{\vector{u},i-1/2}),\cr (\rho
  E)^{n+1}_i = (\rho E)^{n}_i &-&\frac{\Delta t}{\Delta x}((\rho
  E)_{i+1/2}u^*_{i+1/2}+\pi^*_{\vector{u},i+1/2}\cdot\vector{u}^*_{i+1/2}\cr
  &&-(\rho
  E)_{i-1/2}u^*_{i-1/2}-\pi^*_{\vector{u},i-1/2}\cdot\vector{u}^*_{i-1/2}),\cr
  \vector{B}^{n+1}_i = \vector{B}^n_i &-&\frac{\Delta t}{\Delta
    x}(\vector{B}_{i+1/2}u^*_{i+1/2}-B_{x,i+1/2}^{-u^*}\vector{u}^*_{i+1/2}\cr
  &&-\vector{B}_{i-1/2}u^*_{i-1/2}+B_{x,i-1/2}^{-u^*}\vector{u}^*_{i-1/2}).
\end{eqnarray}
The global scheme of stencil 2 is stable under the most restrictive CFL
condition between the magneto-acoustic and transport sub-systems. The scheme of
stencil 1 is stable under a CFL condition involving the sum of the speeds of the
magneto-acoustic and transport subsystem as demonstrated in
\cite{bourgeois:2024} and in Sect.~\ref{sect:entropy}.

\section{Entropy analysis}\label{sect:entropy}

In this section, we first introduce under which conditions the
  1D relaxation solver is entropy-satisfying.
For a non-constant $B_x$ in a multi-dimensional setup, it is clear that the
fully-conservative solver is not entropy-satisfying: on the $-u/L$ wave, $B_x$
is the only quantity that jumps, hence, induces a jump in internal energy because
of the last Riemann invariant in (\ref{eq:invariants}).  Similarly to
\cite{bouchut:2010}, an entropy satisfying solver will require the introduction
of an entropic correction on the induction equation to get a symmetric version of the MHD equations. We
will present the multi-dimensional entropy-satisfying solver at the end of the
section.

The choice of the relaxation parameter $c=c_a=c_b$ for the 3+1 wave approximate
Riemann solver and $c_a$, $c_b$ for the 5+1 wave solver is made to ensure that
the solver is entropy satisfying for a constant $B_x$ in 1D.  If for all
intermediate states $\vector{U}^*_{l,r}$, one has $\tau^*_{l,r}>0$ and
\begin{eqnarray}\label{eq:inequality}
  (\rho^2 c_s^2)_{*,l,r} &\le& c_b^2,\cr \tau^*_{l,r}-\frac{B_x^2}{c_a^2}&\ge&
  0,\cr \left(B_{y,l,r}^{2}+B_{z,l,r}^{2}\right)&\le& (c^2_b-(\rho^2
  c_s^2)_{*,l,r})\left(\tau^*_{l,r}-\frac{B_x^2}{c_a^2}\right),
\end{eqnarray}
with $(\rho^2 c_s^2)_{*,l,r} \equiv \sup_{\rho \in
  (\rho_*,\rho_l,\rho_r)}(\rho^2c_s^2(\rho,s_{l,r}))$, there exists a numerical
flux function $q^n_{i+1/2}= q(\vector{U}^n_i,\vector{U}^n_{i+1})$, consistent
with zero (see \cite{chalons:2014}) such that
\begin{eqnarray}\label{eq:entropy_ineq}
\rho^{n+1}_i s(\vector{U}^{n+1}_i) - \rho_i^n s(\vector{U}^{n}_i)
&+&\frac{\Delta t}{\Delta x}(q_{i+1/2}^n+(\rho s)_{i+1/2}u^*_{i+1/2}\cr
&&-q_{i-1/2}^n-(\rho s)_{i-1/2}u^*_{i-1/2}) \ge 0.
\end{eqnarray}
Following \cite{bouchut:2010}, optimal choices of $c_a$ and $c_b$ for smooth
solutions are given by 
\begin{eqnarray}\label{eq:cacb}
  c_a^2 &=& \rho(B_x^2+|B_x|\sqrt{B_y^2+B_z^2})\cr c_b^2 &=&
  \rho^2c_s^2+\rho(B_y^2+B_z^2+|B_x|\sqrt{B_y^2+B_z^2})
\end{eqnarray}
for the 5+1 wave solver and $c = \rho c_{mf}$ for the 3+1 wave solver.  Optimal
choices for discontinuous solutions are given in \cite{bouchut:2010}, however,
in all the tests performed in Sect.~\ref{sect:tests} the smooth version has been
sufficient to ensure stability and is therefore preferred for its low
computational cost. As noted
in \cite{bouchut:2010}, the diffusion of the 5+1 solver is zero when $B_x=0$ or
$B_y^2+B_z^2=0$ which means that the solver is exact in these conditions. We,
however, point out that this is exactly where the MHD system is not strictly
hyperbolic with $c_{ma}=0$ for $B_x=0$ and $c_{ma}=c_{ms}$ for $B_y^2+B_z^2=0$.
Therefore, in practice, we employ a more diffusing approximation for the choices of $c_a$
and $c_b$ by using the following inequality $|B_x|\sqrt{B_y^2+B_z^2}\le
  (B_x^2+B_{y}^2+B_z^2)/2$: 
\begin{eqnarray}\label{eq:cacb_isotropic}
  c_a^2 &=& \rho(B_x^2+(B_x^2+B_{y}^2+B_z^2)/2)\cr c_b^2 &=&
  \rho^2c_s^2+\rho(B_y^2+B_z^2 + (B_x^2+B_{y}^2+B_z^2)/2)
\end{eqnarray}
to ensure the use of a stable strictly hyperbolic approximation even when $B_x$
or $B_y^2+B_z^2$ vanishes. It also helps with the isotropy of the numerical
diffusion whenever there is a large difference between the normal and transverse
magnetic intensity, avoiding the generation of spurious patterns.  We decompose
the proof of the entropy analysis of the global scheme into an entropy analysis
of each sub-system, magneto-acoustic and transport, respectively.

\subsection{Entropy analysis of the magneto-acoustic sub-system in 1D}

\textbf{Proposition 1: } Let $s_{l,r} = s(\tau_{l,r},e_{l,r})$.  If the
inequality
\begin{equation}\label{eq:e_inequality}
e^*_{l,r}\ge e(\tau^*_{l,r},s_{l,r})
\end{equation}
is verified, there exists a numerical flux function $q^n_{i+1/2}=
q(\vector{U}^n_i,\vector{U}^n_{i+1})$, consistent with zero such that
\begin{equation} \label{eq:mag-ac_ineq}
L^{n+1-}_i \rho^{n+1-}_i s(\tau^{n+1-}_i,e^{n+1-}_i)- \rho_i^n
s(\tau^{n}_i,e^{n}_i) + \frac{\Delta t}{\Delta x}(q_{i+1/2}^n-q_{i-1/2}^n) \ge 0
\end{equation}

\paragraph{Proof} According to Eq. (\ref{eq:thermo}), at  fixed $\tau$, $e(\tau,s)$ is
an increasing function of $s$, hence $e(\tau^*_{l,r},s_{l,r}^*)\ge
e(\tau^*_{l,r},s_{l,r})$ implies $s_{l,r}^* \ge s_{l,r}$. This inequality then
implies that for any $c>0$
\begin{equation}
0\ge -c(s_l^*-s_l) + c(s_r - s_r^*)
\end{equation}
which is consistent with the integral form of the entropy inequality $\partial_t
(s(\tau,e)) \ge 0$. As in \cite{chalons:2014}, this implies the existence of
$q^n_{i+1/2}=q(\vector{U}^n_i,\vector{U}^n_{i+1})$ such that
\begin{equation}
s(\tau^{n+1-}_i,e^{n+1-}_i)- s(\tau^{n}_i,e^{n}_i) + \tau^n_i\frac{\Delta
  t}{\Delta x}(q_{i+1/2}^n-q_{i-1/2}^n) \ge 0
\end{equation}
The inequality (\ref{eq:mag-ac_ineq}) follows from $L^{n+1-}_i \rho^{n+1-}_i
=\rho^n_i$.

\textbf{Proposition 2: } The 5+1 wave approximate Riemann solver associated to
the relaxation (\ref{eq:full_relax}) of the magneto-acoustic sub-system is
positive and satisfies all discrete entropy inequalities whenever for all
intermediate states $\vector{U}^*_{l,r}$, $\tau^*_{l,r}$ are positive and the
inequalities (\ref{eq:inequality}) are verified.

\paragraph{Proof} According to (\ref{eq:invariants}), the  5+1  wave relaxation
Riemann problem has the same Riemann invariants as \cite{bouchut:2010} apart
from the addition of $B_x$ as a strong Riemann invariant of the $-u/L$ wave. $B_x$
has therefore to be understood as evaluated locally according to (\ref{eq:bx}). By
introducing the decomposition into elementary dissipation terms similarly as in
\cite{bouchut:2002}, using the Riemann invariants (\ref{eq:invariants}) and
defining $\sigma(\vector{U})=p(\tau,s=s_{l,r})+\vector{B}^2/2$, one can show
that
\begin{equation}\label{eq:magique}
e(\tau^*_{l,r},s_{l,r}) - e^*_{l,r} = D_0(\vector{U}^*_{l,r},\vector{U}_{l,r}) -
\frac{1}{2}\left|\frac{\sigma(\vector{U}^*_{l,r})
  \vector{n}-B_x\vector{B}^*-\pi_\vector{u}^*}{c_\vector{u}} \right|^2,
\end{equation}
with $D_0$ the dissipation associated to the central wave given by
\begin{eqnarray}
D_0(\vector{U}^*_{l,r},\vector{U}_{l,r})&=&
e(\tau^*_{l,r},s_{l,r})-e(\tau_{l,r},s_{l,r})+
p(\tau^*_{l,r},s_{l,r})\left(\tau^*_{l,r}-\tau_{l,r}\right)\cr &&+\frac{1}{2
  c_b^2}\left(\sigma(\vector{U}^*_{l,r}) -\sigma(\vector{U}_{l,r})\right)^2\cr
&&-\left(
\tau_{l,r}-B_x^2/c_a^2\right)\frac{1}{2}\left|\vector{B}^*-\vector{B}_{l,r}
\right|^2.
\end{eqnarray}
The proof of proposition 2 then follows directly from the entropy analysis of
\cite{bouchut:2007} who showed that under (\ref{eq:inequality}) and by using
\ref{eq:magique}, the inequality (\ref{eq:e_inequality}) is verified.

The final part of the analysis requires to give the conditions under
  which the relaxation approximation is positive for the intermediate states of
  the specific volume  $\tau^*_{l,r}>0$. These conditions for the relaxation
  parameters are provided in proposition 3.3 of \cite{bouchut:2010}, however we
  do not explicitly specify them here because we will use a less restrictive
  choice with Eq.~(\ref{eq:cacb_isotropic}) which seems sufficient in practice in
  all the numerical tests performed in Sect.~\ref{sect:tests}.

\subsection{Entropy analysis of the transport sub-system in 1D}

By using $u^\pm = \frac{u\pm |u|}{2}$, the transport step of the global scheme
of stencil 2 can be written in the form
\begin{equation}
\vector{U}^{n+1}_i = \frac{\Delta t}{\Delta x} u^{*,+}_{i-1/2}
\vector{U}^{n+1-}_{i-1}-\frac{\Delta t}{\Delta x} u^{*,-}_{i+1/2}
\vector{U}^{n+1-}_{i+1} + \left(1-\frac{\Delta t}{\Delta x}
(u^{*,+}_{i-1/2}-u^{*,-}_{i+1/2})\right) \vector{U}^{n+1-}_{i},
\end{equation}
hence $\vector{U}^{n+1}_i $ is a convex combination of
$\vector{U}^{n+1-}_{i-1},\vector{U}^{n+1-}_i$ and $\vector{U}^{n+1-}_{i+1}$ as their
pre-factors are positive and sum to $1$.
By convexity of the function $\vector{U}\rightarrow -\rho s(\vector{U})$
\begin{equation}\label{eq:trans_ineq}
\rho^{n+1}_i s(\vector{U}^{n+1}_i) \ge \rho^{n+1-}_i L^{n+1-}_i
s(\vector{U}^{n+1-}_i) -\frac{\Delta t}{\Delta x}((\rho
s)_{i+1/2}u^*_{i+1/2}-(\rho s)_{i-1/2}u^*_{i-1/2}).
\end{equation}
By combining, the inequalities (\ref{eq:mag-ac_ineq}) and (\ref{eq:trans_ineq}) we
obtain the inequality (\ref{eq:entropy_ineq}). Following
\cite{bourgeois:2024}, the global scheme of stencil 1 can be written in the form
\begin{equation} \label{eq:average}
\vector{U}^{n+1}_i = \alpha_i \vector{U}^A_i +
(1-\alpha_i)\vector{U}^T_i
\end{equation}
for any $\alpha_i\in]0,1[$ and
\begin{eqnarray}
\vector{U}^A_i&=& \vector{U}^{n}_i + \frac{1}{\alpha_i}\frac{\Delta t}{\Delta x}
(\vector{U}^{n+1-}_i L^{n+1-}_i-\vector{U}^{n}_i),\cr \vector{U}^T_i &=&
\vector{U}^{n}_i -\frac{1}{1-\alpha_i}\frac{\Delta t}{\Delta x}
(u^*_{i+1/2}\vector{U}_{i+1/2}-u^*_{i-1/2}\vector{U}_{i-1/2}),
\end{eqnarray}
with $\vector{U}^A_i$ corresponding to a magneto-acoustic update with $\Delta
t^{A} = \frac{1}{\alpha_i}\Delta t$ and $\vector{U}^T_i$ corresponding to a
conservative transport update also with $\Delta t^T = \frac{1}{1-\alpha_i}\Delta
t$.  Following \cite{bourgeois:2024}, $\vector{U}^T_i/\rho^ {n+1}_i$ can
be written as a convex combination of $\vector{U}^n_i/\rho^{n}_i$. Thus, we can
also obtain (\ref{eq:entropy_ineq}) by using the convexity of (\ref{eq:average})
under the CFL conditions:
\begin{eqnarray}
  \max_{i \in \mathbb{Z}} ( (u^*_{i-1/2})^+-(u^*_{i+1/2})^-)
  \frac{1}{\alpha_i}\Delta t \le \Delta x. \\ \max_{i \in \mathbb{Z}}\left(
  \frac{c_{rf,i}}{\rho_i}\right) \frac{1}{1-\alpha_i}\Delta t \le \frac{\Delta
    x}{2}.
\end{eqnarray}
As the local choice of $\alpha_i$ is free, we can pick it so that both
conditions coincide, giving the following condition for the stencil 1 scheme:
\begin{equation}
  \max_{i \in \mathbb{Z}} ( (u^*_{i-1/2})^+-(u^*_{i+1/2})^- +2
  \frac{c_{rf,i}}{\rho_i})\Delta t \le \Delta x.
\end{equation}

\subsection{Entropic correction for multi-dimensional MHD}
Similarly to \cite{bouchut:2010}, we introduce an entropic correction on the
induction equation proportional to $\vector{\nabla}\cdot\vector{B}$,
\begin{equation}\label{eq:noncons}
\partial_t \vector{B} + \vector{\nabla}\cdot(\vector{u}\otimes\vector{B}
-\vector{B}\otimes\vector{u}) + \vector{u} \vector{\nabla}\cdot\vector{B}= 0.
\end{equation}
The rest of the MHD system is not changed and, of course, (\ref{eq:noncons}) is
equivalent to the standard form when $\vector{\nabla}\cdot\vector{B}=0$. Smooth
solutions follow the entropy evolution 
\begin{equation}\label{eq:entropy_scheme_symm}
 \partial_t(\rho s) + \vector{\nabla}\cdot(\rho s\vector{u}) = 0.
\end{equation}
We recall the discretization of the source term as \cite{bouchut:2010}
which results in two different values of $B_x^{-u^*}$ at an interface,
$B_{x,i+1/2,l}^{-u^*}=B_{x,i}^n$ and $B_{x,i+1/2,r}^{-u^*}=B_{x,i+1}^n$, hence
giving a non-conservative discretization of the induction equation with
\begin{eqnarray}\label{eq:powell_st}
\vector{B}^{n+1}_i = \vector{B}^n_i &-&\frac{\Delta t}{\Delta
  x}(\vector{B}_{i+1/2}u^*_{i+1/2}-B_{x,i}^{n}\vector{u}^*_{i+1/2}\cr
&&-\vector{B}_{i-1/2}u^*_{i-1/2}+B_{x,i}^{n}\vector{u}^*_{i-1/2}).
\end{eqnarray}

With this non-conservative source term, the evolution equation of $B_x$ is
simply $\partial_t B_x=0$ and the system becomes symmetric with an additional wave
centered at 0 instead of the $-u/L$ wave \cite{godunov:1972}. The strong Riemann
invariant $B_x$ 
jumps at 0, similarly to the other Riemann invariant (\ref{eq:invariants}). As in
\cite{bouchut:2010}, the 3+1 and 5+1 approximate Riemann solvers with the
non-conservative source term are entropy satisfying with the same proof
presented above, $B_x$ simply needs to be understood as evaluated locally with a
jump on the central wave.

We emphasize that the normal component of the magnetic field for $B_x^{-u^*}$ in
(\ref{eq:powell_st}) is always the value at cell center $B_{x,i}^n$ both at first
and second order. As noted by \cite{klingenberg:2010}, the source term
vanishes for smooth solutions at second order if one uses the reconstructed
values at interfaces.  The proposed discretization in (\ref{eq:powell_st}) avoids
this problem and can be employed for both 1st and 2nd order.

\section{Numerical results}\label{sect:tests}

In order to take advantage of the fully conservative and entropy-satisfying
solvers, we use an hybrid strategy in our simulations by switching between both
depending on the regime of the flow. On
cells where the plasma beta number $\beta = p / \frac{\vector{B}^2}{2}$ is
inferior to a tunable threshold $\beta_{min}$ or where  the local Alfvén number
$Al=\sqrt{\rho}\frac{\mid u \mid}{\mid B \mid}$ is superior to another tunable
treshold ${Al}_{max}$, we use locally the entropy-satisfying solver instead of
the fully conservative solver. In all our experiments,  we set
$\beta_{min}=10^{-3}$ and  ${Al}_{max}=10$. The entropic correction is only
activated in the
specifically designed low-plasma-beta blast problem (see
Sect. \ref{sect:Blast_low_PB}) and the field loop advection test case
(See Sect. \ref{sect:FieldLoop}).

All the simulations performed in this section are using the stencil 1
  solver and a MUSCL-Hancock scheme
\cite{van_leer}, with second order accuracy in space with states
reconstructions at interfaces and second order accuracy in time with a
predictor-corrector step at 
half time-step. We perform the extrapolation on the primitive variables
($\rho, p, \vector{u}, \vector{B}$) and use the classical minmod limiter in
order to ensure the admissibility of the Riemann states. The
time-step is computed with $\Delta t = \mathrm{CFL} \times \Delta x/(c_{mf}+|\vector{u}|)$
with $c_{mf}$ the speed of the fast magneto-acoustic waves. With a standard
MUSCL-Hancock scheme the CFL condition with $\mathrm{CFL<0.5}$ ensures a
positive numerical scheme. In practice, this CFL condition is often observed to
be too restrictive and we use in all simulations a fixed CFL number of
$0.8$. We also use an ideal gas equation of state.  All
numerical experiments were conducted using the one step $5+1$ waves solver with
$c_a$ and $c_b$ given by Eq. (\ref{eq:cacb_isotropic}) to avoid the loss of
hyperbolicity of the relaxation whenever $B_x$ or $B_y$ and $B_z$ vanish.
We also use the $3+1$ waves solver for the 2D rotated shock tube test
  to provide a comparison with the $5+1$ waves solver.




\subsection{1D tests cases}\label{sect:1D_tests}

In this section, we reproduce several 1D Riemann problems that were used in
\cite{bouchut:2010}. The values of the left and right states, the final time, lenght of the
domain and adiabatic indexes are given in table \eqref{eq:tab_riemann}. The simulations were all
performed with $\Delta x =10^{-2}$. The reference solutions were all generated
with the $5+1$ waves solver using $\Delta x = 5\times 10^{-4}$.
\begin{equation*}\label{eq:tab_riemann}
  \begin{array}{|l|cccc|}
    \hline \text {Test case name, }(\gamma,t_{end},L)& \rho & (u,v,w) & \mathrm{p} &
    (B_{x},B_{y},B_{z}) \\ \hline \text{Dai \& Woodward, }(\frac{5}{3},0.2,1.1) & &&&\\ \text {L
      state} & 1.08 &(1.2,0.01,0.5) & 0.95 & (\frac{4}{\sqrt{4
        \pi}},\frac{3.6}{\sqrt{4 \pi}}, \frac{2}{\sqrt{4 \pi}}) \\ \text {R
      state} & 1.0 & (0.0,0.0,0.0) & 1.0 & (\frac{4}{\sqrt{4
        \pi}},\frac{4}{\sqrt{4 \pi}},\frac{2}{\sqrt{4 \pi}})\\ \hline \text{Brio \& Wu
      I, }(2.0,0.2,1.0)& &&&\\ \text {L state} & 1.0 & (0.0,0.0,0.0) & 1.0 &(
    0.65,1.0, 0.0) \\ \text {R state} & 0.125 & (0.0,0.0,0.0) & 0.1 &(0.65,
    -1.0,0.0) \\ \hline \text{Brio \& Wu II, }(2.0,0.012,1.4)& &&&\\ \text {L state} & 1.0 &
    (0.0,0.0,0.0) & 1000.0 & (0.0,1.0,0.0)\\ \text {R state} & 0.125 &
    (0.0,0.0,0.0) & 0.1 & (0.0,-1.0,0.0) \\ \hline \text {Slow rarefaction,
    }(\frac{5}{3},0.2,1.0)& &&&\\ \text {L state} & 1.0 & (0.0,0.0,0.0) & 2.0 &
    (1.0,0.0,0.0) \\ \text {R state} & 0.2 & (1.186,2.967,0.0) & 0.1368 &
    (1.0,1.6405,0.0) \\ \hline \text {Expansion I, }(\frac{5}{3},0.15,1.4)& &&&\\ \text {L
      state} & 1.0 & (-3.1,0.0,0.0) & 0.45 & (0.0,0.5,0.0)\\ \text {R state} &
    1.0 & (3.1,0.0,0.0) & 0.45 & (0.0,0.5,0.0) \\ \hline \text {Expansion II,
    }(\frac{5}{3},0.15,1.4)& &&&\\ \text {L state} & 1.0 & (-3.1,0.0,0.0) & 0.45 &
    (1.0,0.5,0.0)\\ \text {R state} & 1.0 & (-3.1,0.0,0.0) & 0.45 &
    (1.0,0.5,0.0) \\ \hline
    \end{array}
\end{equation*}

\subsubsection{Dai-Woodward shock tube}\label{sect:DW}
\begin{figure}[H]
    \centering
    \includegraphics[width=0.49\linewidth]{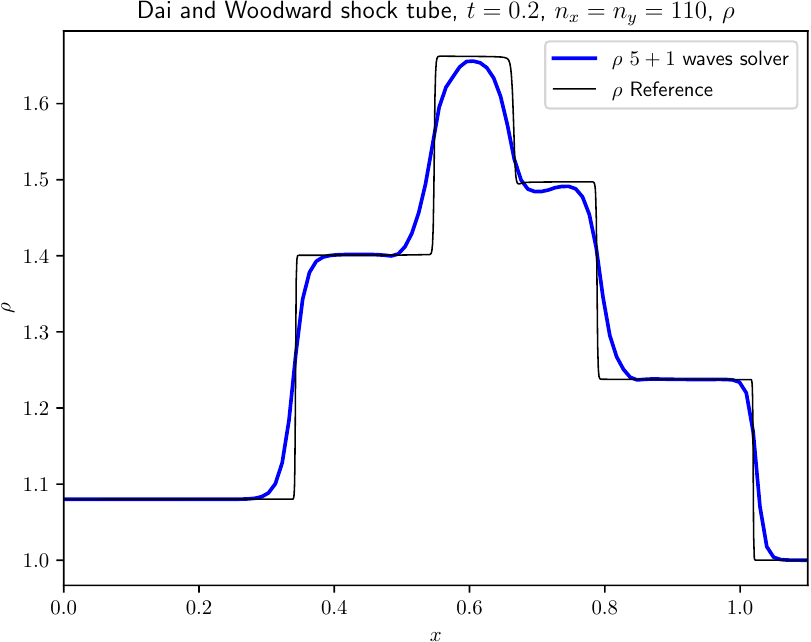}
    \includegraphics[width=0.49\linewidth]{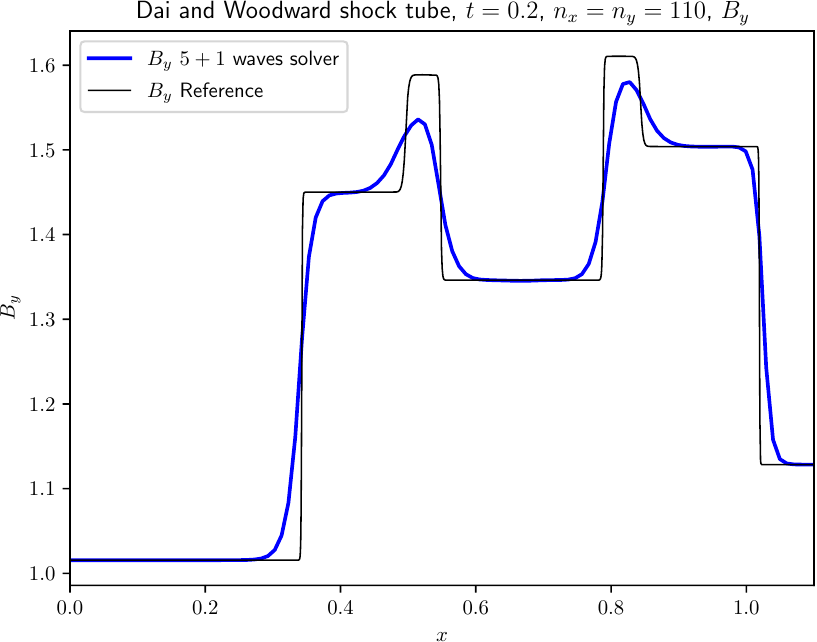}
    \caption{\label{fig:Dai-Woodward_figure} $\rho$ and $B_y$ for the Dai and
      Woodward shock tube at $t=0.2$, $5+1$ waves solver against a reference
      solution.}
\end{figure}

This shock tube configuration was introduced in \cite{DW}. During the
computation, the solution displays the full eigen-structure of the MHD system as
it generates shocks and discontinuities on all fields. We observe in figure
\ref{fig:Dai-Woodward_figure} that our method captures the density and transverse 
magnetic field robustly, without spurious oscillations. We observe the effect of numerical 
diffusion smoothing the various waves. A density undershoot is observed at $x\simeq 0.7$
and is due to the choice of CFL number $0.8$, higher than what the $0.5$ allowed by the stability
analysis of MUSCL methods. These results are very similar to what is obtained in \cite{bouchut:2010}.

\subsubsection{Brio-Wu shock tube, configuration I}\label{sect:BWI}

\begin{figure}[H]
    \centering
    \includegraphics[width=0.49\linewidth]{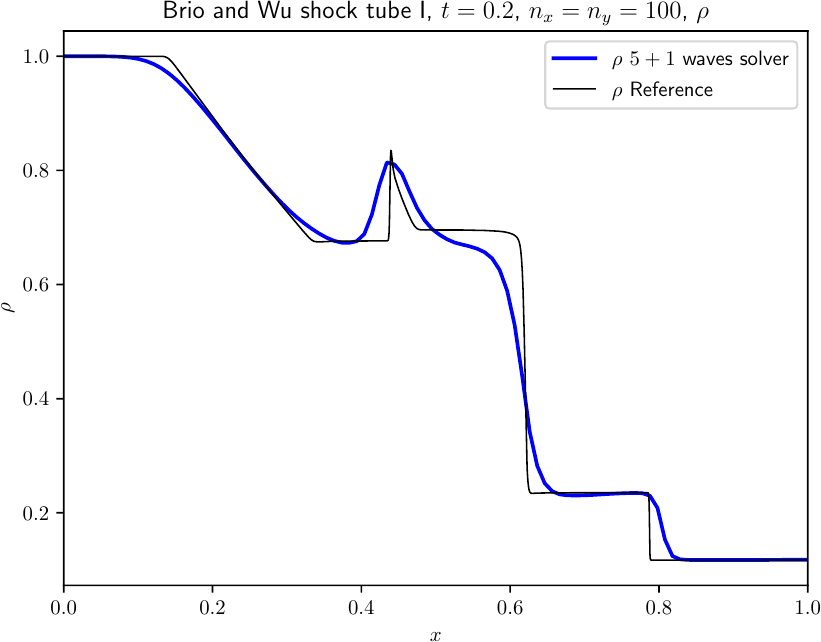}
    \includegraphics[width=0.49\linewidth]{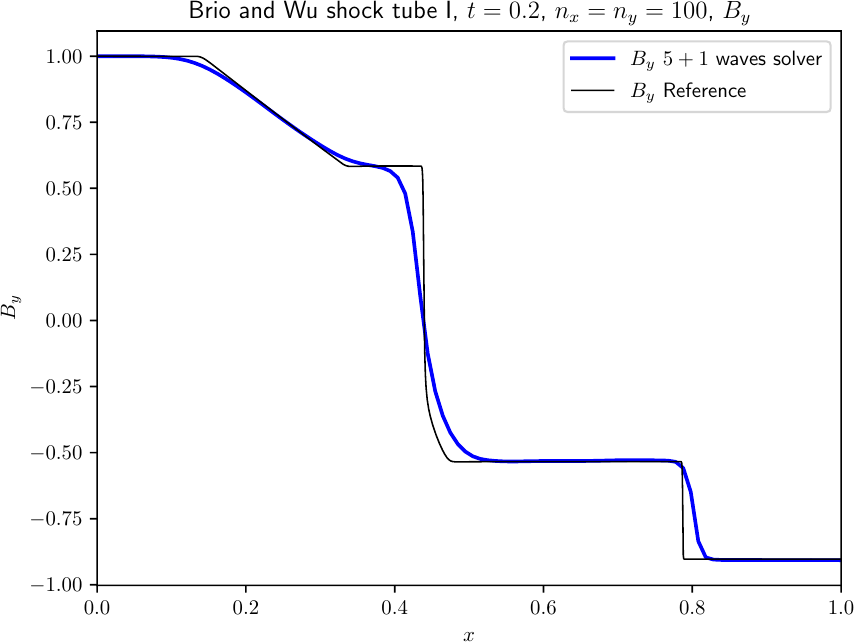}
    \caption{\label{fig:Brio-Wu_I_figure} $\rho$ and $B_y$ for the Brio and Wu
      -I- shock tube at $t=0.2$, $5+1$ waves solver against a reference
      solution.}
\end{figure}

The Brio-Wu shock tube was first introduced in \cite{BRIO1988400}. The solution
of this shock tube is composed of shocks, rarefactions, contact discontinuities
and a compound wave, in this case a discontinuity attached to a slow
rarefaction. In figure \ref{fig:Brio-Wu_I_figure}, we can see that our solver
captures all features of the solution of this Riemann problem. The effect of diffusion 
is mainly observed on the $x\simeq 0.6$ shock and the density peak around
$x\simeq 0.45$ as it is a very fine feature.
 At the same location, the low-resolution result does present a smoothed bump.
These results are very similar to what is obtained in \cite{bouchut:2010}. Note
that as in \cite{bouchut:2010}, the slow shock position does not seem to be well
captured at low resolution.

\subsubsection{Brio-Wu shock tube, configuration II}\label{sect:BWII}
\begin{figure}[H]
    \centering
    \includegraphics[width=0.49\linewidth]{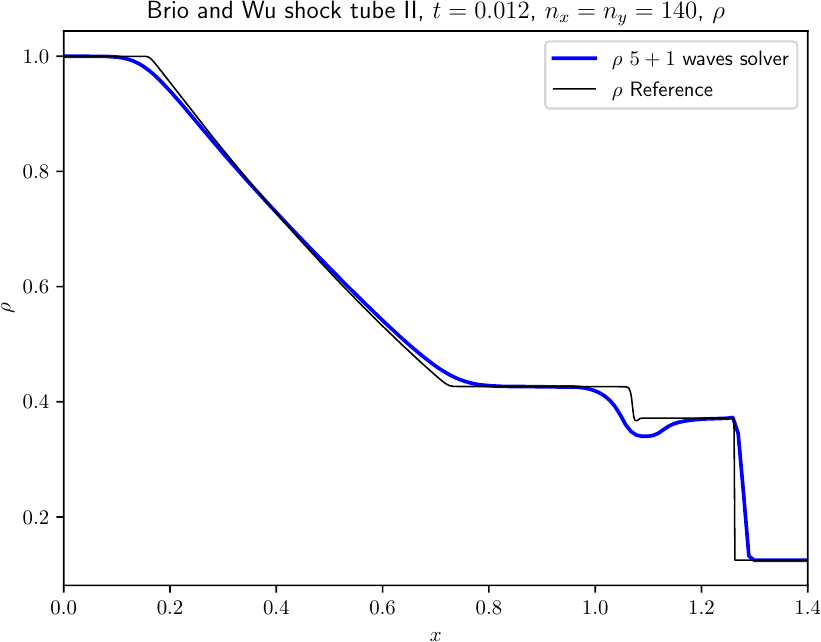}
    \includegraphics[width=0.49\linewidth]{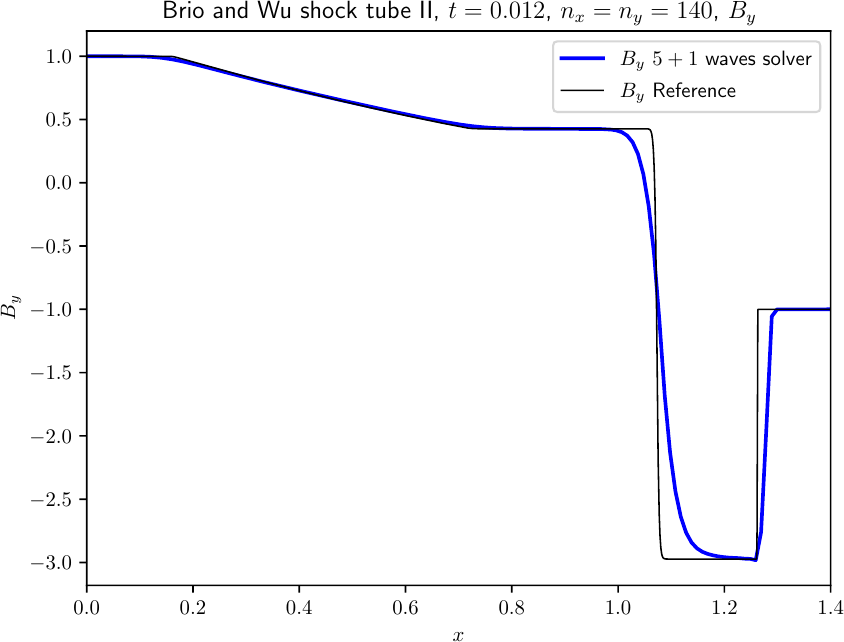}
    \caption{\label{fig:Brio-Wu_II_figure} $\rho$ and $B_y$ for the Brio and Wu
      -II- shock tube at $t=0.012$, $5+1$ waves solver against a reference
      solution.}
\end{figure}

The second Riemann problem from \cite{BRIO1988400} also involves a complex wave
structure but with a high magneto-acoustic Mach number. In figure
\ref{fig:Brio-Wu_II_figure}, we observe that our solver captures all
features of the shock tube, similarly to the results of \cite{bouchut:2010}. The effect 
of diffusion is mainly observed at $x\simeq 1.05$ where a discontinuity and an undershoot 
are observed on the high resolution plot. This corresponds to the smoothed dip observed in the 
low-resolution solution.

\subsubsection{Slow rarefaction tube}\label{sect:SR}
\begin{figure}[H]
    \centering \includegraphics[width=0.49\linewidth]{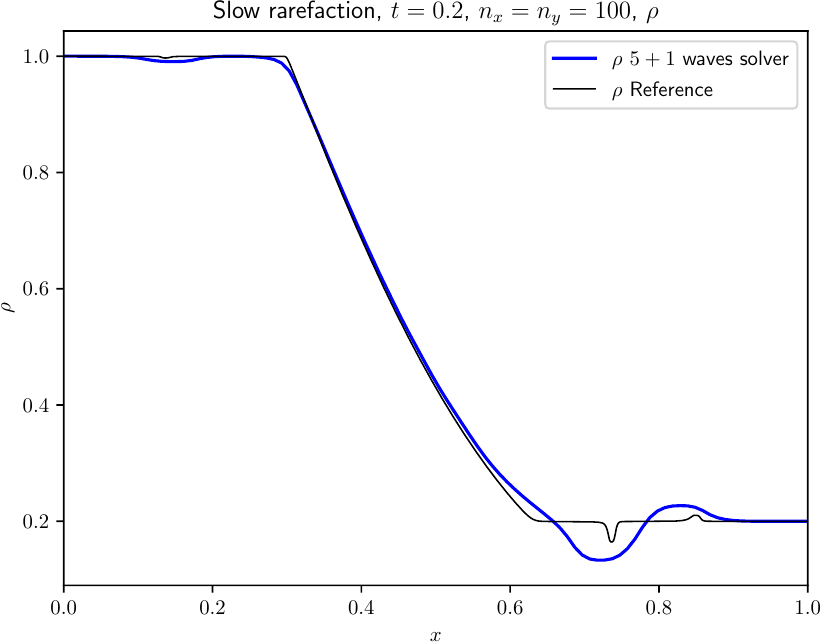}
    \includegraphics[width=0.49\linewidth]{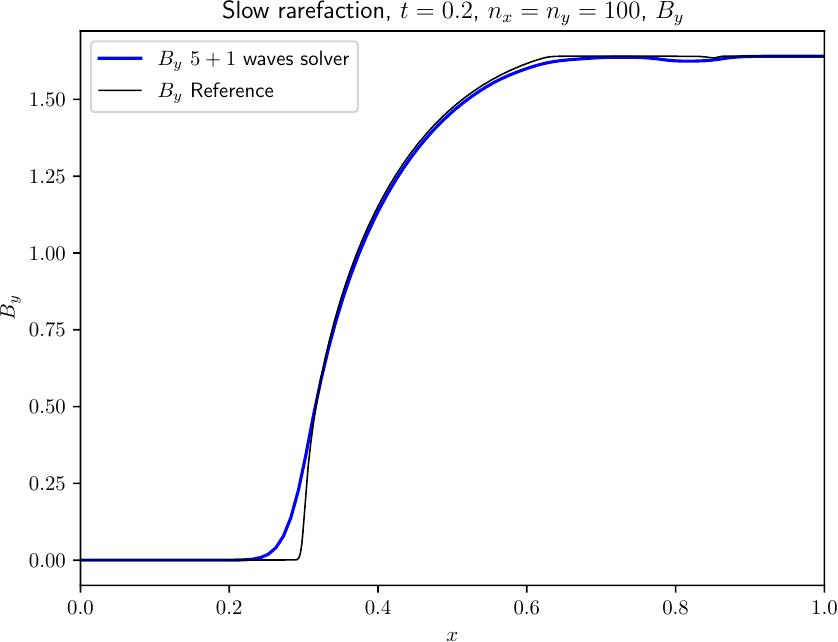}
    \caption{\label{fig:Slow_rarefaction_figure} $\rho$ and $B_y$ for the slow
      rarefaction tube at $t=0.2$, $5+1$ waves solver against a reference
      solution.}
\end{figure}

This test has been first proposed in \cite{Falle}. It involves a sonic point,
where the slow magneto-acoustic speed equals the fluid velocity. This feature is
problematic for linearized method like the Roe solver, but our scheme is stable
as we can see in figure \ref{fig:Slow_rarefaction_figure}, just like the resolution
shown in \cite{bouchut:2010}. The $x\simeq 0.75$ dip and $x\simeq 0.85$ bump present 
on the high-resolution line are smoothed but still present on the low-resolution solution.

\subsubsection{Expansion problem, configuration I}\label{sect:EI}

\begin{figure}[H]
    \centering
    \includegraphics[width=0.49\linewidth]{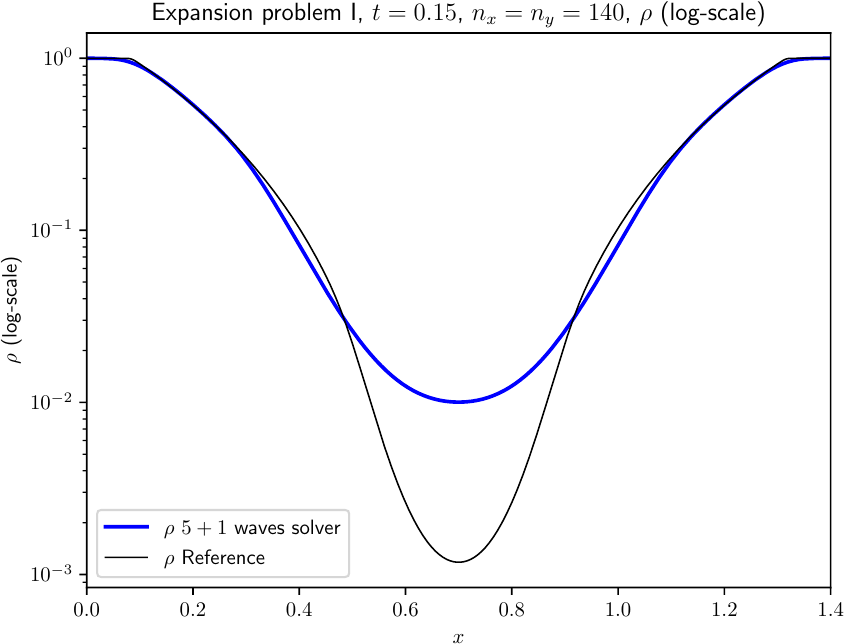}
    \includegraphics[width=0.49\linewidth]{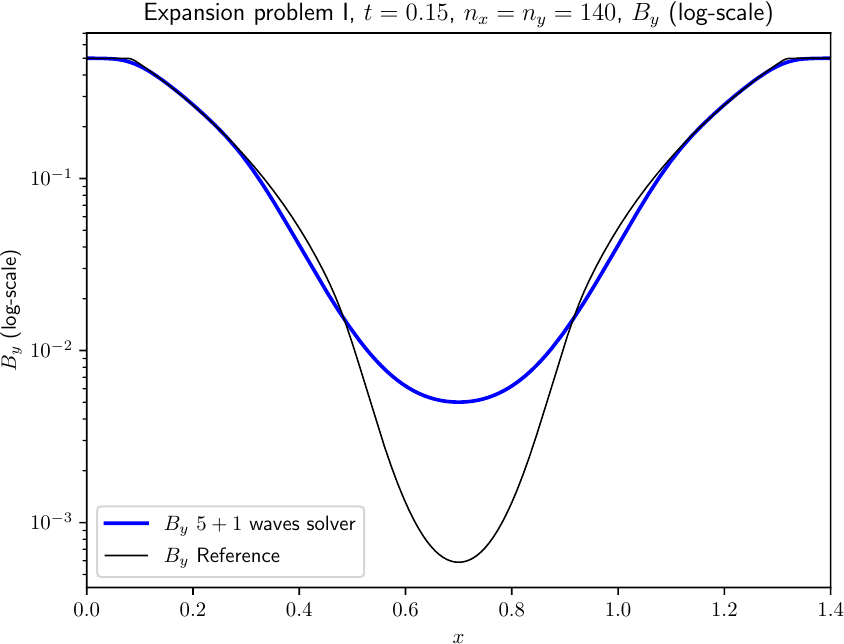}
    \caption{\label{fig:Expansion_I_figure} $\rho$ and $B_y$ for the expansion
      -I- tube at $t=0.15$, $5+1$ waves solver against a reference
      solution. logscale on the $y$-axis.}
\end{figure}

This test is taken from \cite{Miyoshi}. It consists of two out-going rarefaction
separating a low density region that is difficult to tackle in a stable manner. Our
solver is able to simulate this region as we can see in figure
\ref{fig:Expansion_I_figure}. The effect of numerical diffusion on the sharpness
of the $x=0.5$ density and magnetic field dip is visually enhanced by the use of
the log scale. Similar results are found in \cite{bouchut:2010}.

\subsubsection{Expansion problem, configuration II}\label{sect:EII}
\begin{figure}[H]
    \centering
    \includegraphics[width=0.49\linewidth]{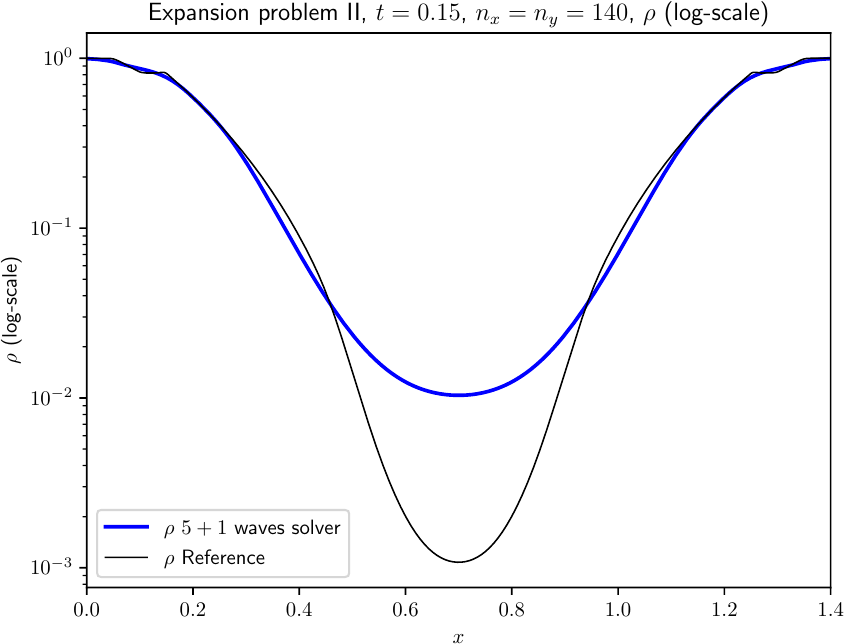}
    \includegraphics[width=0.49\linewidth]{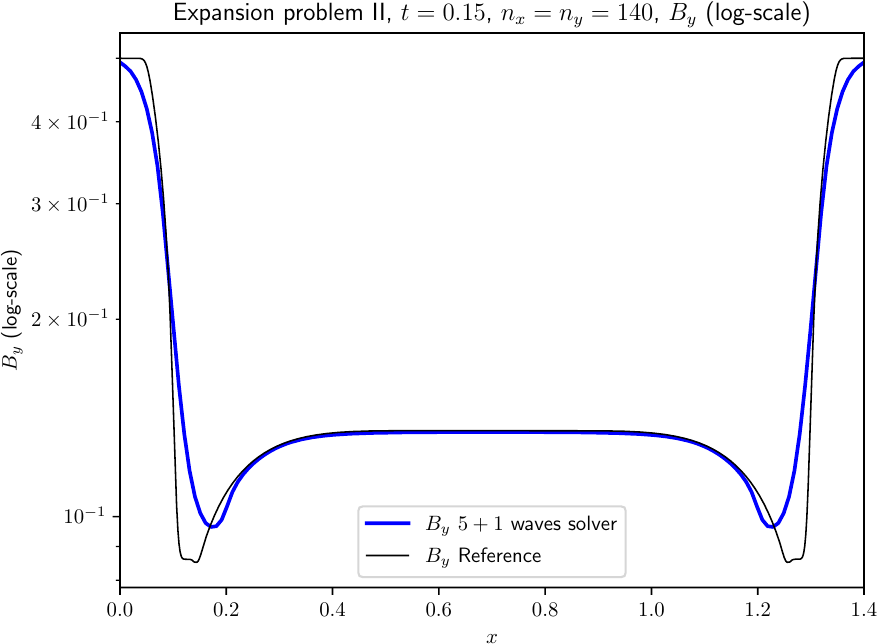}
    \caption{\label{fig:Expansion_II_figure} $\rho$ and $B_y$ for the expansion
      -II- tube at $t=0.15$, $5+1$ waves solver against a reference
      solution. logscale on the $y$-axis.}
\end{figure}

This test is a modification of \ref{sect:EI} suggested by \cite{bouchut:2010}
where we simply set $B_x=1.0$ instead of $0$. Taking $B_x$ nonzero causes the
thermal pressure to be low in the central region which can be hard to tackle
robustly. Nevertheless, we can see in figure \ref{fig:Expansion_II_figure} that
our method is stable and provides results that are very similar to the ones
presented in \cite{bouchut:2010}.

\subsection{2D tests cases}

All 2D test cases are using $\Delta x=\Delta y = \frac{1}{256}$.  We also have
tested all the resolutions between 64 and 2048 without any issue to report. In
all 2D setups, the quantity $r$ always refers to the distance from the
center of the domain.

\subsubsection{Orszag-Tang vortex}\label{sect:Otang}
\begin{figure}[H]
    \centering \includegraphics[width=\linewidth]{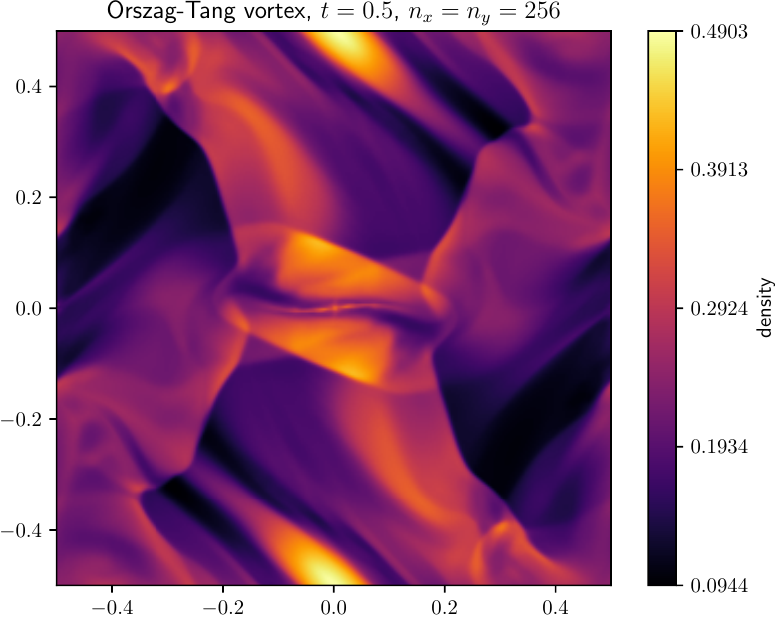}
    \caption{\label{fig:Otang_vortex} Density map of the Orszag-Tang vortex at
      $t=0.5$s}
\end{figure}

The Orszag-Tang vortex test case was first introduced by \cite{orszag_tang_1979}
and has become a standard multi-dimensional benchmark case for ideal MHD. The
dynamic of this vortex involves the formation of shocks as well as interactions
between them which are challenging to simulate robustly.  For instance, 1D
solvers like HLLD straightforwardly extended to 2D fail at this task. We recall
that this problem takes place in the $[0:1]^2$ periodic domain with initial
data:
\begin{eqnarray*}
    \rho(x,y) &=& 25/36\pi,\cr p(x,y) &=& 5/12\pi,\cr \vector{u}(x,y)
    &=& \begin{pmatrix} -\text{sin} \ 2\pi y \\ \text{sin} \ 2 \pi x
    \end{pmatrix},\cr
    \vector{B}(x,y)&=& 1/\sqrt{4\pi} \begin{pmatrix} -\text{sin} \ 2\pi y
      \\ \text{sin} \ 4 \pi x
\end{pmatrix},\cr
    \gamma &=& 5/3.
\end{eqnarray*}
We show the density map at $t=0.5$ in Figure \ref{fig:Otang_vortex}. We observe
that the shocks and discontinuities are well captured without spurious
numerical artifacts. We also notice the usual "eye-shape" high frequency feature
at the center of the domain, demonstrating the accuracy of our solver. Note that
this test does not show any low $\beta$ zone. Thus, the solver is fully
conservative with respect to $\vector{B}$ as the entropic correction is never
activated.

\subsubsection{Rotated shock tube}\label{sect:Rotated_shock}

\begin{figure}[H]
    \centering
    \includegraphics[width=0.65\linewidth]{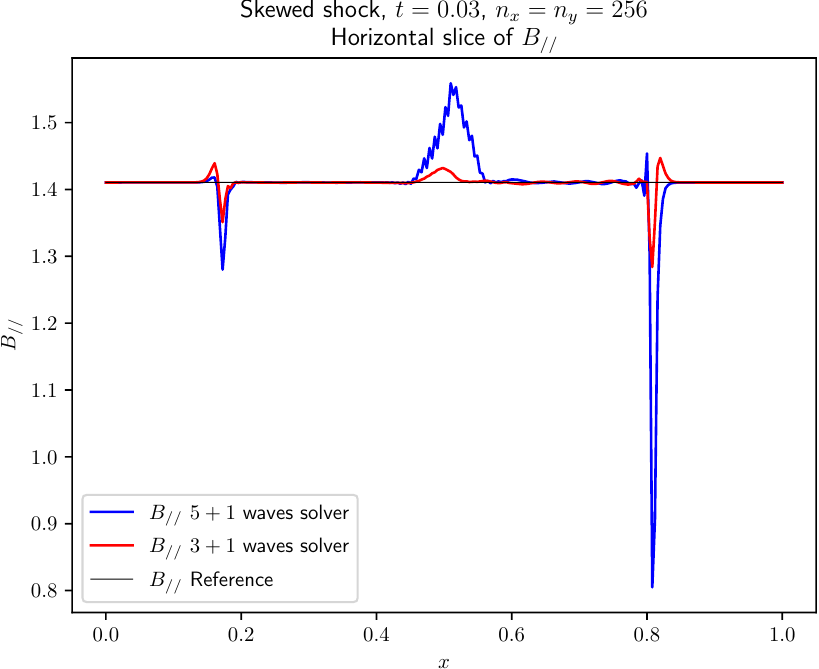}
    \caption{\label{fig:Rotated_shock_figure} Parallel component of the magnetic
      field along the rotated shock propagation at t=0.03.}
\end{figure}

The rotated shock tube problem has been proposed in \cite{TOTH2000605}.  It
consists of a 1D shock tube rotated by an angle $\theta$ in order to obtain a 2D
shock propagation that is not aligned with the grid. The test takes place in the
$[0:1]^2$ square with Neumann boundary conditions. The setup is given by:
\begin{eqnarray*}
  \theta &=& \text{arctan}(-2),\cr \vector{R}(\theta)& =& \begin{pmatrix}
    \text{sin}\ \theta&\text{cos}\ \theta\\ \text{cos}\ \theta&
    -\text{sin}\ \theta
    \end{pmatrix},\cr
   \vector{u}_0 &=&\begin{pmatrix} 0\\ 10
   \end{pmatrix}, \cr
   \vector{B_0}&=&\frac{5}{\sqrt{4\pi}} \begin{pmatrix} 1\\ 1
\end{pmatrix},\cr
(x_\theta,y_\theta) &= & (\tan \theta (x-0.5),y - 0.5),
\end{eqnarray*}

\begin{eqnarray*}
  \rho(x,y)&=& 1, \cr \vector{B}(x,y) &=& \vector{R}(\theta)\vector{B_0},\cr
  \vector{u}(x,y) & =&\left\{\begin{array}{ll} \vector{R}(\theta)\vector{u_0}&
  \text { for } x_\theta<y_\theta, \\ -\vector{R}(\theta)\vector{u_0} & \text {
    elsewhere}.
    \end{array}\right. . \\
        p(x,y) &=&\left\{\begin{array}{ll} 20& \text { for } x_\theta<y_\theta,
        \\ 1 & \text { elsewhere}.
        \end{array}\right. . \\
\end{eqnarray*}
Note that the magnetic field is initialized as a constant on the whole domain,
hence the condition $\nabla \cdot \vector{B}=0$ is verified at the beginning of
the computation.  Our solver is able to robustly and accurately simulate this
rotated shock propagation.  A quantity of interest in this problem is the
component of the magnetic field that is parallel to the shock
propagation. Without discretization error, this quantity should remain constant
similarly to $B_x$ in a purely 1D setup.  In figure
\ref{fig:Rotated_shock_figure}, we show the component of the magnetic field that
is parallel to the shock propagation, with both $3+1$ and $5+1$ solvers. Both
schemes produces discretization errors at the location of discontinuities, the
errors with the $5+1$ waves solver are larger than the errors with the $3+1$
waves solver.  These errors can be compared with \cite{TOTH2000605} for
constrained transport schemes and we point out that the $3+1$ and $5+1$ waves
solvers produce less oscillations around the discontinuities.

\subsubsection{MHD Blast - standard configuration}\label{sect:Blast}
\begin{figure}[H]
    \centering \includegraphics[width=\linewidth]{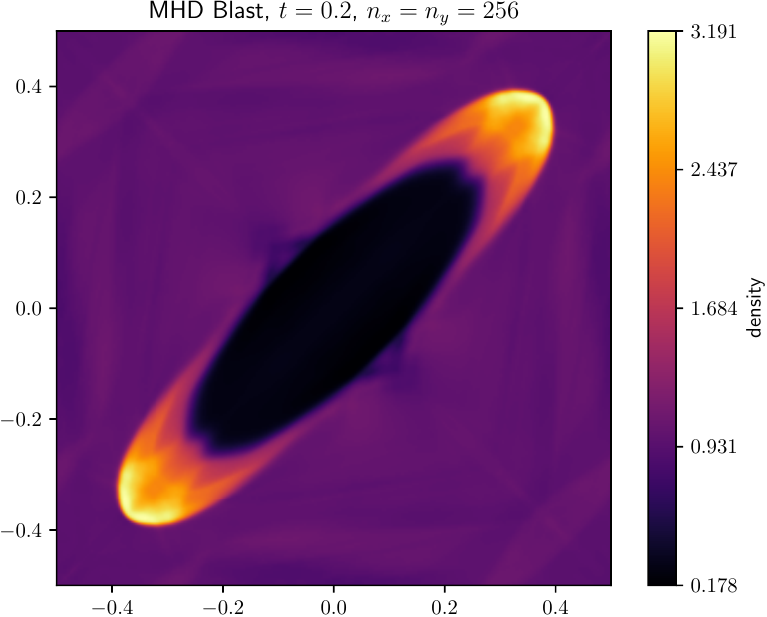}
    \caption{\label{fig:MHD_Blast_figure} Density map of the MHD Blast at
      $t=0.2$s}.
\end{figure}

The Blast test case was introduced in \cite{STONE2009139}. The setup takes place
in the periodic $[0:1]^2$ square.  A circular region of radius $r_{c}=0.1$ is
initialized with a greater pression than the rest of the domain.  As the
computation starts, the blast expands outwards in an elliptical shape due to the
presence of a magnetic field.  We recall the exact setup:
\begin{eqnarray*}
      p(x, y) &=&\left\{\begin{array}{ll} 10& \text { for } r<r_c, \\ 0.1 &
      \text { for } r\geq r_c,
     \end{array}\right. \cr
     \vector{B}(x,y)&=& \begin{pmatrix} \sqrt{2\pi} \\ \sqrt{2\pi}
     \end{pmatrix},\cr
    \gamma &=& 5/3, \cr \rho(x,y) &=& 1, \cr \vector{u}(x,y) &=&0.
 \end{eqnarray*}
Our numerical method is able to simulate the expansion of this blast wave
accurately and is stable as demonstrated in figure \ref{fig:MHD_Blast_figure}
where we show the density map at $t=0.2$.  We can see that the expanding wave is
well captured. Note that this test does not show any low $\beta$ zone. Thus,
the solver is fully conservative with respect to $\vector{B}$.

\subsubsection{MHD blast -  Low  $\beta$ configuration}\label{sect:Blast_low_PB}

\begin{figure}[H]
  \centering \includegraphics[width=\linewidth]{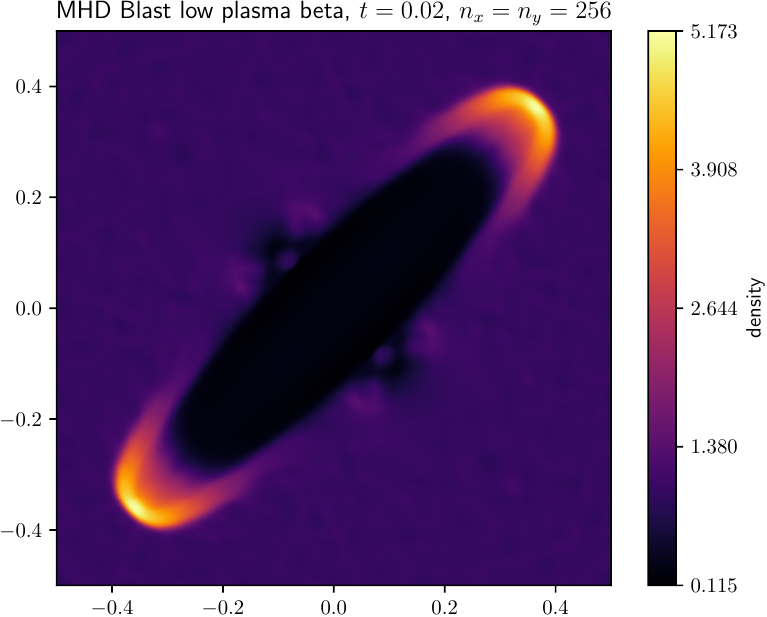}
  \includegraphics[width=\linewidth]{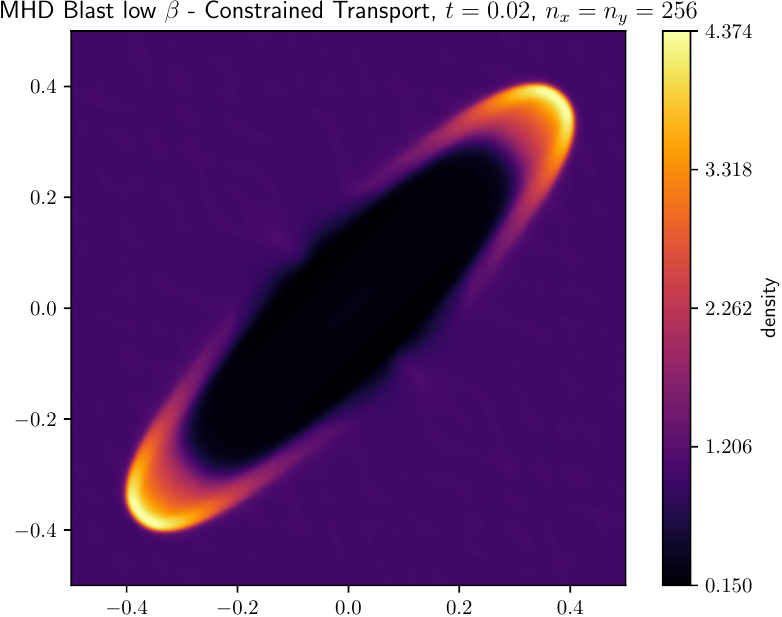}
  \caption{\label{fig:MHD_Blast_low_plasma_beta_figure} Density map of the low
    $\beta$ MHD blast at $t=0.2$s with our solver and the Heracles code's
    constrained transport method \cite{gonzalez:2007}, \cite{vides:2013}.}
\end{figure}

This test case is inspired from \cite{BALSARA20127504}. It consists of the same
setup as section \ref{sect:Blast} with a lower $\beta \simeq 10^{-6}$:
\begin{eqnarray*}
  p(x, y) &=&\left\{\begin{array}{ll} 1000& \text { for } r<r_c, \\ 0.1 & \text
  { for } r\geq r_c,
    \end{array}\right. \cr
      \vector{B}(x,y)&=& \begin{pmatrix} 250/\sqrt{2} \\ 250/\sqrt{2}
    \end{pmatrix},\cr
      \gamma &=& 1.4, \cr \rho(x,y) &=& 1,\cr \vector{u}(x,y) &=& 0.
\end{eqnarray*}
The dynamic of the low $\beta$ blast wave is the same as in \ref{sect:Blast} but
is harder to tackle as the simulation reaches the limit of the admissibility
domain ($e\simeq 0$) and develops strong $\vector{B}$ gradients.  Note that the
$5+1$ wave solver and the constrained transport method \cite{vides:2013} fail to
 produce an admissible result as the computation presents negative internal
energies (directly after few iterations).  We point out that the $5+1$ solver
seems, however, more robust than the constrained transport method on such
problems: for lower values of the magnetic field $25/\sqrt{2}$, the relaxation
solver is stable while the constrained transport method fails after few
iterations.  It is possible to still get an admissible result by artificially
forcing the internal energy to stay above a small threshold (hence loosing
energy conservation), a solution used here with the constrained transport
method, or by using an entropic correction on the induction equation (hence
loosing the magnetic field
conservation), a solution used here with the $5+1$ waves relaxation solver. In
figures \ref{fig:MHD_Blast_low_plasma_beta_figure}, we show the density map of
this test case at $t=0.02$ with our method and the energy-fixed constrained
transport solver from the Heracles code \cite{gonzalez:2007}.  Both methods are
able to capture the low $\beta$ Blast propagation, however, we point out that
the $5+1$ waves solver is less
diffusing as it reaches higher values for the magnetic field up (+18\%).

\subsubsection{MHD Rotor}\label{sect:Rotor}
The MHD Rotor test case was first introduced in \cite{BALSARA1999270}. The setup
consists of launching a rapidly spinning cylinder in a light ambient fluid. This
rotation sends strong torsional Alfvén waves in the surrounding fluid.  We
initialize the solution in the $[0:1]^2$ periodic square as following:

 \begin{eqnarray*}
    p(x, y)&=& 1.0, \cr \rho(x, y)& =&\left\{\begin{array}{ll} 10& \text { for }
    r<r_0, \\ 1+9f & \text { for } r\geq r_1 \ \& \ r\leq r_0,\\ 1 & \text {
      elsewhere}
    \end{array}\right. \cr
\vector{u}(x, y) &=&\left\{\begin{array}{ll} \frac{u_0}{r_0}\ (0.5-y, x-0.5) &
\text { for } r<r_0, \\ \frac{f\ u_0}{r_0}\ (0.5-y, x-0.5) & \text { for } r\geq
r_1 \ \& \ r\leq r_0,\\ (0,0) & \text { elsewhere}
\end{array}\right. \cr
      \vector{B}(x,y)&=& \begin{pmatrix} 5/\sqrt{4\pi} \\ 0
    \end{pmatrix},\\
    \gamma &=& 1.4, \cr (r_0,r_1) &=& (0.1,0.115),\cr f&=& (r_1-r)/(r_1-r_0),\cr
    u_0&=&2.
\end{eqnarray*}

\begin{figure}[H]
    \centering \includegraphics[width=\linewidth]{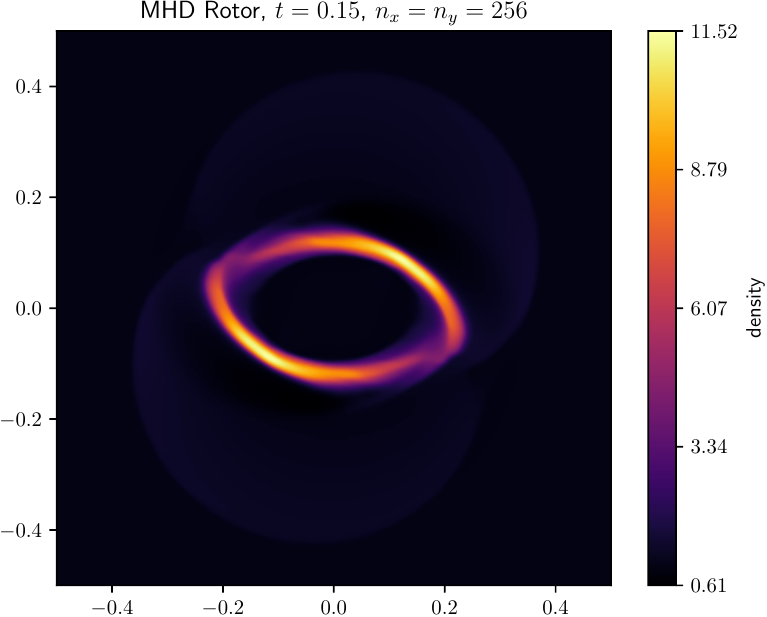}
    \caption{\label{fig:MHD_Rotor_figure} Density map of the MHD Rotor at
      $t=0.15$s}
\end{figure}

We show the result of our simulation in figure \ref{fig:MHD_Rotor_figure}. We
observe that the central shear ring as well as the torsional waves are well captured
 by our solver. Note that this simulation does not require
the use of entropic correction.  Thus, the solver is fully conservative with
respect to $\vector{B}$.

\subsubsection{Field loop advection}\label{sect:FieldLoop}

\begin{figure}[H]
  \centering \includegraphics[width=\linewidth]{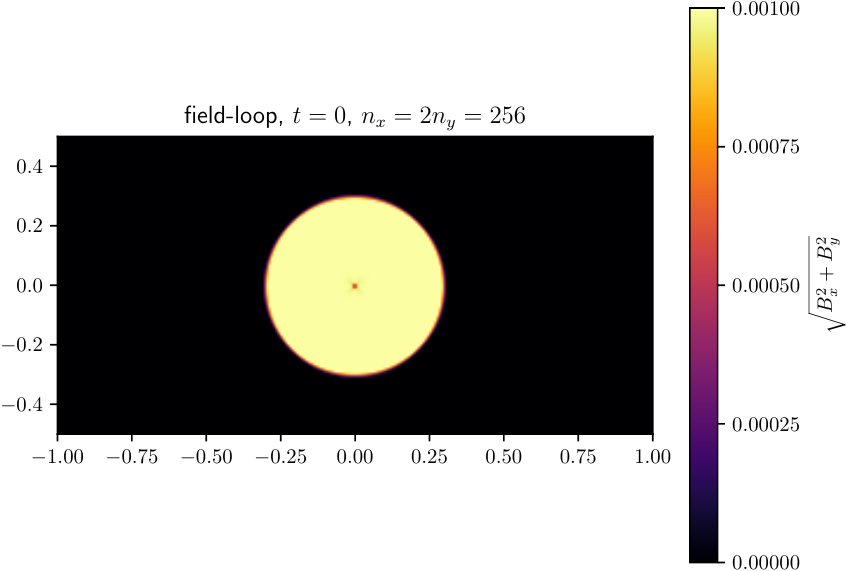}
  \caption{\label{fig:field_loopIC} Magnetic intensity of the field loop
    advection at time $t=0$.}
\end{figure}

\begin{figure}[H]
  \centering \includegraphics[width=\linewidth]{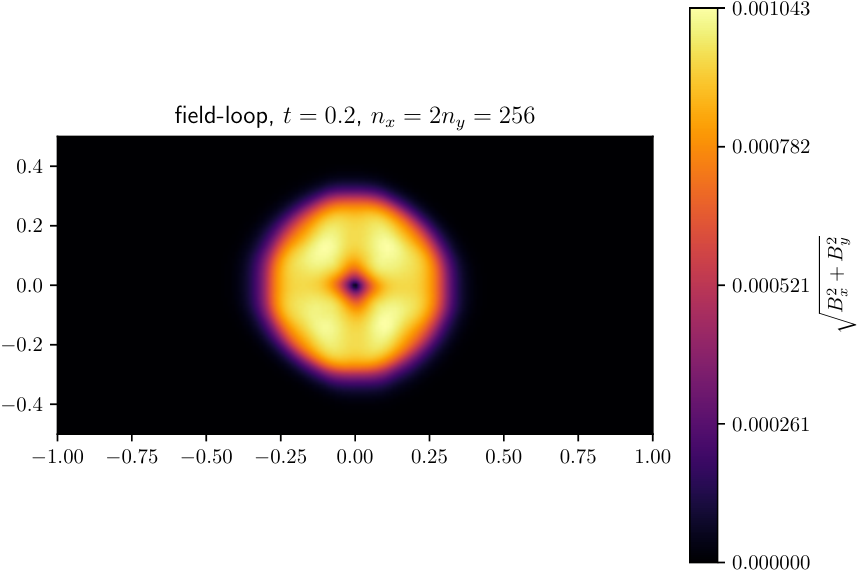} \centering
  \includegraphics[width=\linewidth]{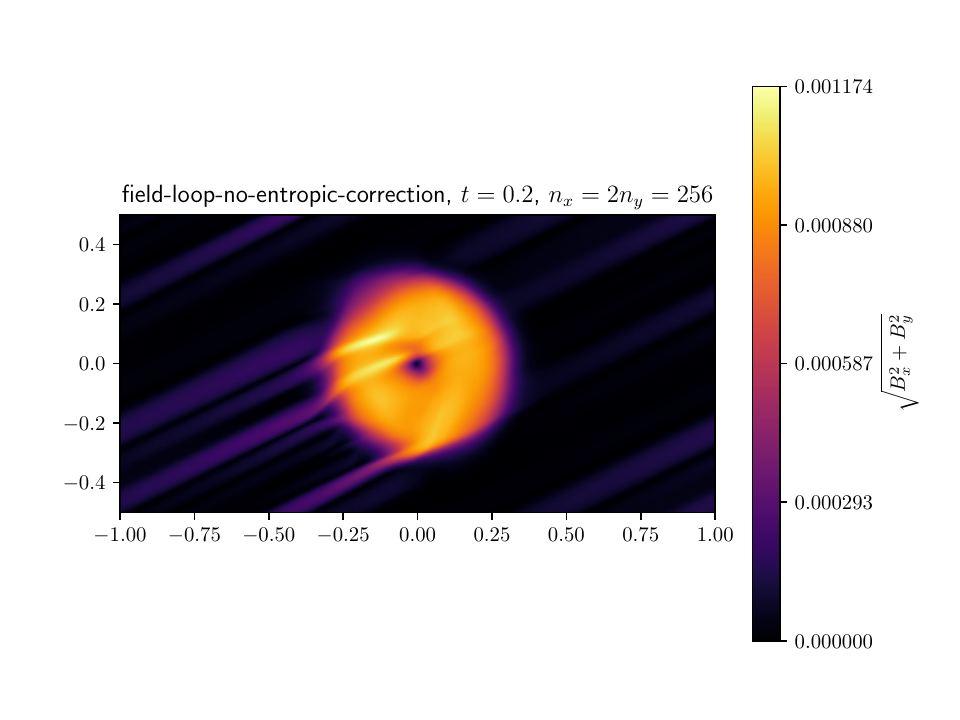}
  \caption{\label{fig:field_loop} Magnetic intensity of the field loop advection
    at time $t=2.0$. Top: With the entropic correction. Bottom: Without the
    entropic correction.}
\end{figure}

  This test was introduced in \cite{toth1996comparison} and involves advecting a
  field loop (a cylindrical current distribution) diagonally across the
  grid. One can choose any arbitrary angle. For the 2D results presented here,
  the problem domain is defined as $-1<x<1$ and $-0.5<y<0.5$. The flow has an
  inclination with $V_x=2$ and $V_y=1$. Both the density and pressure are set to
  $1.0$, with the gas constant given by $\gamma = 5/3$. Periodic boundary
  conditions are applied across the domain.  The magnetic field is initialized
  using an arbitrary vector potential.  We set $A_z = \max([A_0 ( r_0 - r
    )],0)$. This results in $(B_x, B_y)(r)=\frac{A_0}{r}(-x, y)$ if $r<r_0$, and
  $(0,0)$ otherwise. We chose $A_0=0.001$ and set the radius for the loop as
  $r_0 = 0.3$. After a duration of $t=2.0s$, the field loop is expected to have
  been advected and returned to its initial state.  The quality of the solution
  can be assessed by comparing it to the initial solution shown in figure
  \ref{fig:field_loopIC}. The magnetic intensity, defined as
  $I=\sqrt{B_x^2+B_y^2}$, obtained with our $5+1$ waves solver, is illustrated
  in figure \ref{fig:field_loop}. One can observe that the entropic correction
  helps with preserving the shape of the cylinder and suppresses the spurious
  patterns observed with the conservative method. The source terms are activated
  here as the Alfvén number is above ${Al}_{max}=10$ in this test.

\section{Conclusion and discussion}

In this paper, we have developed a new multi-dimensional, robust, and
cell-centered finite-volume solver for ideal MHD.  The solver is based on
splitting and relaxation techniques, and can easily be extended to higher
orders because of its reduced stencil. A symmetric version of the solver
has been developed by introducing an entropic correction on the
induction equation, in order to obtain an
entropy-satisfying (but non-conservative for the magnetic field) scheme robust
in low plasma beta regions and accurate in high Alfvén number regions. An other
solution could be to use a floor value for the internal energy as classically
done with constrained transport or divergence cleaning schemes that are not
entropy satisfying. We, however, point out that the fully conservative
relaxation solver is observed to be more robust than constrained transport
schemes on low plasma beta test cases.

This cell-centered scheme could be
coupled to a divergence cleaning or constrained transport method. We, however,
highlight that all the tests we have performed do not seem to require a specific
treatment of the divergence of the magnetic field, and a divergence consistent
with zero with errors proportional to $\Delta x$ and $\Delta t$ at the power of
the order of the spatial and temporal reconstructions seem sufficient. It is a
common belief
that the stability of MHD numerical schemes is closely tied to errors in
magnetic field divergence. However, our research, as presented in this paper,
suggests that this may not always be the case. To illustrate, we have
successfully designed an entropy-satisfying MHD solver using the symmetric form of
MHD equations without specifically
addressing divergence issues. Furthermore, we have found that constrained
transport schemes, while maintaining zero divergence at machine precision, do
not necessarily satisfy entropy conditions and can fail to maintain positive
internal energy in areas of low plasma beta.

Additionally, there is a prevalent view that errors in magnetic divergence
significantly impact the physical accuracy of simulations, potentially leading
to artificial magnetic monopoles. We offer several arguments to challenge this
perspective. Even in constrained transport schemes, certain terms
involving divergence in the conservative forms of the Lorentz force and the
energy evolution equation do not achieve zero at machine precision, despite a zero
divergence. These residual terms in the entropy evolution equation are indeed
the reason why constrained transport schemes are not entropy satisfying.
Moreover, it can be demonstrated that constrained transport schemes
are not immune to divergence errors. For example, the rotated shock tube case
detailed in \cite{TOTH2000605} shows that at the continuous level, a zero divergence
equates to a constant magnetic field parallel to the shock tube. However,
constrained transport schemes do not maintain this constant magnetic field at
machine precision, thus resulting in ``divergence errors'' that are significant
for the physics at play.

In conclusion, while ensuring zero magnetic divergence at machine precision in
simulations is physically relevant, this is only feasible when aligning the grid
to a specific magnetic field configuration. This issue is akin to preserving
angular momentum in a rotating structure, achievable at machine precision only
in a polar grid. Consequently, for simulations with highly dynamic magnetic
fields, maintaining zero divergence at machine precision on a Cartesian grid may
not be as critical with a solver that is entropy-satisfying.

The MHD relaxation solver presented in this paper is a direct extension of 1D
relaxation solvers already used for the Euler equations and can be implemented
in a one-step flux-update algorithm, that can easily be extended to higher
orders and to non-ideal MHD. Because of its simplicity, this solver
should also have improved performances compared to other multi-dimensional MHD
solvers (constrained transport and divergence cleaning) and offers interesting
possibilities for large-scale physical applications on the next generation of
exascale supercomputers.

\appendix

\section{Deriving the conservative MHD equations}\label{sect:nonconstocons}

  \subsection{Useful vector identities}
  \subsubsection{Lorentz's force in conservative form}
  The first identity we derive is 
  \begin{equation}\label{eq:lorentzcons}
  \bm{j}\times \bm{B}=-(\nabla \cdot \bm{B})\bm{B} - \nabla\cdot\left(\frac{\bm{B}^2}{2}\bm{I}  - \bm{B}\otimes \bm{B}\right).
  \end{equation}
  We only verify this equality for the $x$ component as the relationship for the two other components are checked 
  by rotational invariance. We have $\bm{j}\times \bm{B}=(\nabla \times \bm{B})\times \bm{B}$. Expanding the first component, 
  we get $[(\nabla \times \bm{B})\times \bm{B}]_x=B_z\left(\partial_z{B_x}-\partial_x{B_z}\right)-B_y\left(\partial_x{B_y}-\partial_y{B_x}\right)$.
  Moreover, $[\nabla \cdot \left(\frac{\bm{B}^2}{2}\bm{I}\right)]_x=B_x\partial_xB_x +B_y\partial_xB_y +B_z\partial_xB_z$. Lastly, 
  $[\nabla\cdot\left(\bm{B}\otimes \bm{B}\right)]_x= (\nabla\cdot \bm{B})B_x + B_x\partial_xB_x +B_y\partial_yB_x +B_z\partial_zB_x$.
  Collecting the right hand side terms, we get $- (\nabla\cdot \bm{B})B_x - B_x\partial_xB_x -B_y\partial_xB_y -B_z\partial_xB_z+(\nabla\cdot \bm{B})B_x + B_x\partial_xB_x +B_y\partial_yB_x +B_z\partial_zB_x$ 
  where both
  the terms proportional to the divergence of $\bm{B}$ and $B_x\partial_xB_x$ cancel out and provide the desired result. 
  
  \subsubsection{Fully developed Lorentz force}
  Using $\nabla \cdot (\bm{B}\otimes \bm{B}) = \bm{B}(\nabla \cdot \bm{B}) + (\bm{B}\cdot\nabla)\bm{B}$, we get:

  \begin{equation}\label{eq:lorentzdev}
    \bm{j}\times \bm{B}=(\bm{B}\cdot\nabla)\bm{B} - \nabla\left(\frac{\bm{B}^2}{2}\right)
  \end{equation}
  
  \subsubsection{Curl of a cross product}
  \begin{equation}\label{eq:curlcross}
  \nabla\times(\bm{u}\times\bm{B})=\nabla \cdot(\bm{B}\otimes \bm{u} - \bm{u}\otimes \bm{B}).
  \end{equation}
  
  \begin{equation}\label{eq:curlcross2}
    \nabla\times(\bm{u}\times\bm{B})=\bm{u}(\nabla\cdot\bm{B}) - \bm{B}(\nabla \cdot \bm{u}) + (\bm{B}\cdot\nabla)\bm{u}-(\bm{u}\cdot \nabla)\bm{B}
  \end{equation}
  
  \subsubsection{Transport of a squared quantity}
  
  \begin{equation}\label{eq:trcarre}
    ((\bm{u}\cdot \nabla)\bm{A})\cdot \bm{A} = (\bm{u}\cdot \nabla)\frac{\bm{A}^2}{2} = \nabla\left(\frac{\bm{A}^2}{2}\right)\cdot \bm{u}
  \end{equation}
  
  \subsection{Full system}
  
  Our goal is now to go from the non conservative MHD system:
  
  \begin{eqnarray}\label{eq:mhd_nonconsapdx}
    \partial_t \rho + \vector{\nabla}\cdot(\rho\vector{u}) = 0, \cr
    \partial_t(\rho \vector{u}) +\vector{\nabla}\cdot(\rho
    \vector{u}\otimes\vector{u} ) = -\vector{\nabla}p + \vector{j} \times
    \vector{B},\cr \partial_t(\rho e) +
    \vector{\nabla}\cdot(\rho e\vector{u}) = -p \nabla\cdot \bm{u},\cr
    \partial_t \vector{B} - \nabla \times(\bm{u}\times \bm{B})  = 0.
  \end{eqnarray}
  
  to the conservative MHD system.
  \begin{eqnarray}\label{eq:mhd_consapdx}
    \partial_t \rho + \vector{\nabla}\cdot(\rho\vector{u}) = 0, \cr
    \partial_t(\rho \vector{u}) +\vector{\nabla}\cdot(\rho
    \vector{u}\otimes\vector{u} +\vector{\sigma} -\vector{B}\otimes\vector{B} ) =
    0,\cr \partial_t(\rho E)+\vector{\nabla}\cdot(\rho
    E\vector{u}+\sigma\vector{u}- (\vector{B}\cdot\vector{u})\vector{B}) =
    0,\cr \partial_t \vector{B} + \vector{\nabla}\cdot(\vector{u}\otimes\vector{B}
    -\vector{B}\otimes\vector{u})= 0.
  \end{eqnarray}

  Where $e_{mag} =\frac{\bm{B}^2}{2\rho}$ and $\sigma = p+ \frac{\bm{B}^2}{2}$ Obtaining
  the conservative momentum equation is straightforward using \eqref{eq:lorentzcons}, 
  substituting for 
  $\vector{j} \times\vector{B}$ and assuming $\nabla\cdot \bm{B}=0$. 
  Obtaining the conservative induction equation is also straightforward using 
   \eqref{eq:curlcross} (note that using the $\nabla \cdot \bm{B}=0$ hypothesis is not 
   necessary to obtain the induction equation). This leaves us with deriving the 
   total energy equation.
  
  \subsubsection{Kinetic energy evolution equation}
  
   From the non conservative momentum equation, we can deduce the evolution equation of the velocity $\partial_t \bm{u} + (\bm{u}\cdot \nabla)\bm{u}+\bm{\nabla p}/\rho = \bm{j}\times \bm{B}/\rho$. Dotting this 
   equation against $\bm{u}$, we get $\partial_t\left(\frac{\bm{u}^2}{2}\right) + \left((\bm{u}\cdot \nabla)\bm{u}\right)\cdot \bm{u} +\bm{\nabla p}\cdot \bm{u}/\rho = (\bm{j}\times \bm{B})\cdot \bm{u}/\rho$.  Using \eqref{eq:trcarre}, we have
   that $\left((\bm{u}\cdot \nabla)\bm{u}\right)\cdot \bm{u}=(\bm{u}\cdot \nabla)\left(\frac{\bm{u}^2}{2}\right)$. Substituting this transport term and multiplying by $\rho$, we get 
   $\rho \partial_t\left(\frac{\bm{u}^2}{2}\right) + \rho (\bm{u}\cdot \nabla)\left(\frac{\bm{u}^2}{2}\right) +\bm{\nabla p}\cdot \bm{u} = (\bm{j}\times \bm{B})\cdot \bm{u}$.
    Adding $\frac{\bm{u}^2}{2}\left(\partial_t \rho + \nabla \cdot(\rho\bm{u})\right)=0$, we get:
   $\partial_t (\rho \frac{\bm{u}^2}{2}) + \rho (\bm{u}\cdot \nabla)\left(\frac{\bm{u}^2}{2}\right) +\frac{\bm{u}^2}{2}\nabla \cdot(\rho \bm{u})+\bm{\nabla p}\cdot \bm{u} = (\bm{j}\times \bm{B})\cdot \bm{u}$. Since 
   $\rho (\bm{u}\cdot \nabla)\left(\frac{\bm{u}^2}{2}\right) +\frac{\bm{u}^2}{2}\nabla \cdot(\rho \bm{u})=\nabla\cdot\left(\frac{\rho \bm{u}^2 \bm{u}}{2}\right)$, noting $e_{kin}=\frac{\bm{u^2}}{2}$, we get:
  
   \begin{equation}
    \partial_t (\rho e_{kin}) + \nabla\cdot\left(\rho e_{kin} \bm{u}\right) + \bm{\nabla p}\cdot \bm{u} = (\bm{j}\times \bm{B})\cdot \bm{u}.
   \end{equation}
  
   Summing this with the internal energy evolution equation, we get:
   $\partial_t (\rho (e+e_{kin})) + \nabla\cdot\left(\rho (e+e_{kin}) \bm{u} + p\bm{u}\right) = (\bm{j}\times \bm{B})\cdot \bm{u}$. Replacing the right hand side 
   using \eqref{eq:lorentzdev}, we get:

   \begin{equation}\label{eq:evolekinint}
  \partial_t (\rho (e+e_{kin})) + \nabla\cdot\left(\rho (e+e_{kin}) \bm{u} + p\bm{u}\right) = ((\bm{B}\cdot\nabla)\bm{B})\cdot \bm{u} - \nabla\left(\frac{\bm{B}^2}{2}\right)\cdot \bm{u}.
   \end{equation}
    
   \subsubsection{Magnetic energy evolution equation}
  
   Using the identity \ref{eq:curlcross2}, we get $\partial_t \bm{B} -\bm{u}(\nabla\cdot\bm{B}) + \bm{B}(\nabla \cdot \bm{u}) - (\bm{B}\cdot\nabla)\bm{u}+(\bm{u}\cdot \nabla)\bm{B}=0$.
   Dotting against $\bm{B}$, we get 
   
   \begin{equation}\label{eq:evolemag}
   \partial_t (\rho e_{mag})  -(\nabla\cdot\bm{B})(\bm{u}\cdot\bm{B}) + (\nabla \cdot \bm{u})\bm{B}^2 - ((\bm{B}\cdot\nabla)\bm{u})\cdot\bm{B} + ((\bm{u}\cdot \nabla)\bm{B})\cdot\bm{B}=0.
   \end{equation}
  
  \subsubsection{Total energy evolution equation}
  Summing \eqref{eq:evolekinint} and \eqref{eq:evolemag}, we get 
  $\partial_t (\rho E) + \nabla\cdot\left(\rho (e+e_{kin}) \bm{u} + p\bm{u}\right) = ((\bm{B}\cdot\nabla)\bm{B})\cdot \bm{u} - \nabla\left(\frac{\bm{B}^2}{2}\right)\cdot \bm{u}  +(\nabla\cdot\bm{B})(\bm{u}\cdot\bm{B}) - (\nabla \cdot \bm{u})\bm{B}^2 + ((\bm{B}\cdot\nabla)\bm{u})\cdot\bm{B} - ((\bm{u}\cdot \nabla)\bm{B})\cdot\bm{B}$.
  Using \eqref{eq:trcarre}, we have $- ((\bm{u}\cdot \nabla)\bm{B})\cdot\bm{B}  - \nabla\left(\frac{\bm{B}^2}{2}\right)\cdot \bm{u}=\nabla(\bm{B}^2)\cdot \bm{u}$. Moreover,
  Since $\nabla(\bm{B}^2)\cdot \bm{u} + (\nabla \cdot \bm{u})\bm{B^2}= \nabla \cdot(\bm{B}^2 \ \bm{u})=\nabla\cdot(\rho e_{mag} \bm{u} + \bm{B}^2/2 \ \bm{u})$, we can show that:
  \begin{equation}
    \partial_t (\rho E) + \nabla\cdot\left(\rho E \bm{u} + \sigma \bm{u}\right) = ((\bm{B}\cdot\nabla)\bm{B})\cdot \bm{u} +(\nabla\cdot\bm{B})(\bm{u}\cdot\bm{B}) + ((\bm{B}\cdot\nabla)\bm{u})\cdot\bm{B}.
  \end{equation}
  As $((\bm{B}\cdot\nabla)\bm{B})\cdot \bm{u} + ((\bm{B}\cdot\nabla)\bm{u})\cdot \bm{B} = (\bm{B}\cdot\nabla)(\bm{u}\cdot\bm{B}) = \nabla(\bm{B}\cdot \bm{u})\cdot \bm{B}$ and $ \nabla(\bm{B}\cdot \bm{u})\cdot \bm{B}+(\nabla\cdot\bm{B})(\bm{u}\cdot\bm{B})=\nabla\cdot((\nabla\cdot\bm{B})\cdot \bm{B})$, we get the desired result. Note that 
  it is not required to assume $\nabla \cdot \bm{B}=0$ to obtain the conservative total energy equation.
  
  \subsection{Entropy inequality}\label{sect:entropyderiv}
  
  \subsubsection{Entropy inequality of the non conservative MHD system}
  
  We start with the classical result of the entropy inequality of the MHD system \eqref{eq:mhd_nonconsapdx}, starting from 
  the evolution equation of the internal energy.
  We note $D_t=\partial_t + u\nabla$. We have $D_t e= -p(\nabla\cdot \bm{u})\tau$ where $\tau=1/\rho$. From the density evolution equation,
  we have that $D_t \tau = \tau(\nabla\cdot \bm{u})$. Therefore, $D_t e + pD_t \tau =0$. Using the first principle of thermodynamics $de+pd\tau=Tds$, we get
  \begin{equation}
    D_t s =0.
  \end{equation}
  
  \subsubsection{Entropy inequality of the conservative MHD system}
  To go from the non conservative system to the conservative system, we only had to cancel one term in the momentum equation, using the $\nabla \cdot \bm{B}=0$ hypothesis.
  This means that if we are discretizing the conservative momentum equation and that the numerical value of the divergence is not zero, were are in fact 
  discretizing $\partial_t(\rho \vector{u}) +\vector{\nabla}\cdot(\rho
  \vector{u}\otimes\vector{u} ) = -\vector{\nabla}p + \vector{j} \times
  \vector{B} + (\nabla \cdot \bm{B})\bm{B}$. We want to derive the corresponding internal energy equation. We dot the momentum equation against $\bm{u}$ 
  and subtract it to the conservative total energy equation. Doing this, we get $\partial_t(\rho e) +
  \vector{\nabla}\cdot(\rho e\vector{u}) = -p \nabla\cdot \bm{u} -(\nabla \cdot \bm{B})(\bm{B}\cdot \bm{u})$. Performing the same steps as above, we get 
  $D_t e+p D_t \tau = -\tau (\nabla \cdot \bm{B})(\bm{B}\cdot \bm{u})$ thus:
  
  \begin{equation}
    D_t s =-\frac{\tau}{T}(\nabla \cdot \bm{B})(\bm{B}\cdot \bm{u})
  \end{equation}

\section*{Acknowledgements}
P. Tremblin and S. Bulteau would like to acknowledge and thank the ERC for
funding this work under the Horizon 2020 program project ATMO (ID: 757858).

\bibliography{main}

\begin{thebibliography}{10}
\expandafter\ifx\csname url\endcsname\relax
  \def\url#1{\texttt{#1}}\fi
\expandafter\ifx\csname urlprefix\endcsname\relax\def\urlprefix{URL }\fi
\expandafter\ifx\csname href\endcsname\relax
  \def\href#1#2{#2} \def\path#1{#1}\fi

\bibitem{brackbill:1980}
J.~U. {Brackbill}, D.~C. {Barnes}, {The Effect of Nonzero {\ensuremath{\nabla}}
  {\textperiodcentered} B on the numerical solution of the magnetohydrodynamic
  equations}, Journal of Computational Physics 35~(3) (1980) 426--430.
\newblock \href {http://dx.doi.org/10.1016/0021-9991(80)90079-0}
  {\path{doi:10.1016/0021-9991(80)90079-0}}.

\bibitem{ryu:1998}
D.~{Ryu}, F.~{Miniati}, T.~W. {Jones}, A.~{Frank}, {A Divergence-free Upwind
  Code for Multidimensional Magnetohydrodynamic Flows}, ApJ 509~(1) (1998)
  244--255.
\newblock \href {http://arxiv.org/abs/astro-ph/9807228}
  {\path{arXiv:astro-ph/9807228}}, \href {http://dx.doi.org/10.1086/306481}
  {\path{doi:10.1086/306481}}.

\bibitem{dai:1998}
W.~{Dai}, P.~R. {Woodward}, {On the Divergence-free Condition and Conservation
  Laws in Numerical Simulations for Supersonic Magnetohydrodynamical Flows},
  ApJ 494~(1) (1998) 317--335.
\newblock \href {http://dx.doi.org/10.1086/305176} {\path{doi:10.1086/305176}}.

\bibitem{dedner:2001}
A.~Dedner, F.~Kemm, D.~Kr\"{o}ner, C.~Munz, T.~Schnitzer, M.~Wessenberg,
  {Hyperbolic Divergence Cleaning for the MHD Equations}, Journal of
  Computational Physics 175 (2002) 645--673.

\bibitem{evans:1988}
C.~R. {Evans}, J.~F. {Hawley}, {Simulation of Magnetohydrodynamic Flows: A
  Constrained Transport Model}, ApJ 332 (1988) 659.
\newblock \href {http://dx.doi.org/10.1086/166684} {\path{doi:10.1086/166684}}.

\bibitem{balsara:1999}
D.~S. {Balsara}, D.~S. {Spicer}, {A Staggered Mesh Algorithm Using High Order
  Godunov Fluxes to Ensure Solenoidal Magnetic Fields in Magnetohydrodynamic
  Simulations}, Journal of Computational Physics 149~(2) (1999) 270--292.
\newblock \href {http://dx.doi.org/10.1006/jcph.1998.6153}
  {\path{doi:10.1006/jcph.1998.6153}}.

\bibitem{TOTH2000605}
G.~Tóth,
  \href{https://www.sciencedirect.com/science/article/pii/S0021999100965197}{The
  $\nabla \cdot b = 0$ constraint in shock-capturing magnetohydrodynamics
  codes}, Journal of Computational Physics 161~(2) (2000) 605--652.
\newblock \href {http://dx.doi.org/https://doi.org/10.1006/jcph.2000.6519}
  {\path{doi:https://doi.org/10.1006/jcph.2000.6519}}.
\newline\urlprefix\url{https://www.sciencedirect.com/science/article/pii/S0021999100965197}

\bibitem{fromang:2006}
S.~{Fromang}, P.~{Hennebelle}, R.~{Teyssier}, {A high order Godunov scheme with
  constrained transport and adaptive mesh refinement for astrophysical
  magnetohydrodynamics}, A\&A 457~(2) (2006) 371--384.
\newblock \href {http://arxiv.org/abs/astro-ph/0607230}
  {\path{arXiv:astro-ph/0607230}}, \href
  {http://dx.doi.org/10.1051/0004-6361:20065371}
  {\path{doi:10.1051/0004-6361:20065371}}.

\bibitem{gallice:2003}
G.~Gallice, {Positive and Entropy Stable Godunov-Type Schemes for Gas Dynamics
  and MHD Equations in Lagrangian or Eulerian Coordinates}, Numer. Math. 94~(4)
  (2003) 673--713.

\bibitem{bouchut:2007}
F.~Bouchut, C.~Klingenberg, K.~Waagan, A multiwave approximate riemann solver
  for ideal {MHD} based on relaxation. i: theoretical framework, Numerische
  Mathematik 108~(1) (2007) 7--42.

\bibitem{bouchut:2010}
F.~Bouchut, C.~Klingenberg, K.~Waagan, A multiwave approximate riemann solver
  for ideal {MHD} based on relaxation {II}: numerical implementation with 3 and
  5 waves, Numerische Mathematik 115~(4) (2010) 647--679.

\bibitem{godunov:1972}
S.~Godunov, Symmetric form of the magnetohydrodynamic equation, Tech. rep.,
  Computer Center, Novosibirsk, USSR (1972).

\bibitem{busto:2023}
S.~Busto, M.~Dumbser, \href{https://doi.org/10.1137/22M147815X}{A new
  thermodynamically compatible finite volume scheme for magnetohydrodynamics},
  SIAM Journal on Numerical Analysis 61~(1) (2023) 343--364.
\newblock \href {http://arxiv.org/abs/https://doi.org/10.1137/22M147815X}
  {\path{arXiv:https://doi.org/10.1137/22M147815X}}, \href
  {http://dx.doi.org/10.1137/22M147815X} {\path{doi:10.1137/22M147815X}}.
\newline\urlprefix\url{https://doi.org/10.1137/22M147815X}

\bibitem{chalons:2014}
C.~Chalons, M.~Girardin, S.~Kokh,
  \href{https://www.cambridge.org/core/product/identifier/S1815240616000748/type/journal_article}{An
  all-regime {Lagrange}-projection-like scheme for the gas dynamics equations
  on unstructured meshes}, Comm. in Comp. Phys. 20~(1) (2016) pp.~188--233.
\newblock \href {http://dx.doi.org/10.4208/cicp.260614.061115a}
  {\path{doi:10.4208/cicp.260614.061115a}}.
\newline\urlprefix\url{https://www.cambridge.org/core/product/identifier/S1815240616000748/type/journal_article}

\bibitem{padioleau:2019}
T.~Padioleau, P.~Tremblin, E.~Audit, P.~Kestener, S.~Kokh,
  \href{https://dx.doi.org/10.3847/1538-4357/ab0f2c}{A high-performance and
  portable all-mach regime flow solver code with well-balanced gravity.
  application to compressible convection}, The Astrophysical Journal 875~(2)
  (2019) 128.
\newblock \href {http://dx.doi.org/10.3847/1538-4357/ab0f2c}
  {\path{doi:10.3847/1538-4357/ab0f2c}}.
\newline\urlprefix\url{https://dx.doi.org/10.3847/1538-4357/ab0f2c}

\bibitem{bourgeois:2024}
R.~Bourgeois, P.~Tremblin, S.~Kokh, T.~Padioleau,
  \href{https://www.sciencedirect.com/science/article/pii/S0021999123006897}{Recasting
  an operator splitting solver into a standard finite volume flux-based
  algorithm. the case of a lagrange-projection-type method for gas dynamics},
  Journal of Computational Physics 496 (2024) 112594.
\newblock \href {http://dx.doi.org/https://doi.org/10.1016/j.jcp.2023.112594}
  {\path{doi:https://doi.org/10.1016/j.jcp.2023.112594}}.
\newline\urlprefix\url{https://www.sciencedirect.com/science/article/pii/S0021999123006897}

\bibitem{hirt:1974}
C.~Hirt, A.~Amsden, J.~Cook,
  \href{https://www.sciencedirect.com/science/article/pii/0021999174900515}{An
  arbitrary lagrangian-eulerian computing method for all flow speeds}, Journal
  of Computational Physics 14~(3) (1974) 227--253.
\newblock \href
  {http://dx.doi.org/https://doi.org/10.1016/0021-9991(74)90051-5}
  {\path{doi:https://doi.org/10.1016/0021-9991(74)90051-5}}.
\newline\urlprefix\url{https://www.sciencedirect.com/science/article/pii/0021999174900515}

\bibitem{godlewski1996}
E.~Godlewski, P.~Raviart,
  \href{https://books.google.fr/books?id=9BwMIDMmTmcC}{Numerical Approximation
  of Hyperbolic Systems of Conservation Laws}, no. 118 in Applied Mathematical
  Sciences, Springer, 1996.
\newline\urlprefix\url{https://books.google.fr/books?id=9BwMIDMmTmcC}

\bibitem{despres:2010}
B.~Despr{\'e}s, \href{https://books.google.sn/books?id=yVbnQbqt1JcC}{Lois de
  Conservations Eul{\'e}riennes, Lagrangiennes et M{\'e}thodes Num{\'e}riques},
  Math{\'e}matiques et Applications, Springer Berlin Heidelberg, 2010.
\newline\urlprefix\url{https://books.google.sn/books?id=yVbnQbqt1JcC}

\bibitem{jin:1995}
S.~Jin, Z.~Xin, \href{https://api.semanticscholar.org/CorpusID:13245844}{The
  relaxation schemes for systems of conservation laws in arbitrary space
  dimensions}, Communications on Pure and Applied Mathematics 48 (1995)
  235--276.
\newline\urlprefix\url{https://api.semanticscholar.org/CorpusID:13245844}

\bibitem{suliciu:1998}
I.~Suliciu, \href{https://api.semanticscholar.org/CorpusID:121171038}{On the
  thermodynamics of rate-type fluids and phase transitions. i. rate-type
  fluids}, International Journal of Engineering Science 36 (1998) 921--947.
\newline\urlprefix\url{https://api.semanticscholar.org/CorpusID:121171038}

\bibitem{coquel:2001}
F.~Coquel, E.~Godlewski, B.~Perthame, A.~In, P.~Rascle,
  \href{https://api.semanticscholar.org/CorpusID:115535063}{Some new godunov
  and relaxation methods for two-phase flow problems}, 2001.
\newline\urlprefix\url{https://api.semanticscholar.org/CorpusID:115535063}

\bibitem{bouchut:2004}
F.~Bouchut, Nonlinear stability of finite Volume Methods for hyperbolic
  conservation laws: And Well-Balanced schemes for sources, Springer Science \&
  Business Media, 2004.

\bibitem{chalons:2008}
C.~Chalons, J.-F. Coulombel,
  \href{https://www.sciencedirect.com/science/article/pii/S0022247X08007099}{Relaxation
  approximation of the euler equations}, Journal of Mathematical Analysis and
  Applications 348~(2) (2008) 872--893.
\newblock \href {http://dx.doi.org/https://doi.org/10.1016/j.jmaa.2008.07.034}
  {\path{doi:https://doi.org/10.1016/j.jmaa.2008.07.034}}.
\newline\urlprefix\url{https://www.sciencedirect.com/science/article/pii/S0022247X08007099}

\bibitem{coquel:2010}
F.~Coquel, Q.~L. Nguyen, M.~Postel, Q.~H. Tran, Entropy-satisfying relaxation
  method with large time-steps for euler ibvps, Math. Comp. 79 (2010)
  1493--1533.

\bibitem{despre:2011}
B.~Despr\'{e}s, \href{https://doi.org/10.1142/S0219891611002329}{A new
  lagrangian formulation of ideal magnetohydrodynamics}, Journal of Hyperbolic
  Differential Equations 08~(01) (2011) 21--35.
\newblock \href
  {http://arxiv.org/abs/https://doi.org/10.1142/S0219891611002329}
  {\path{arXiv:https://doi.org/10.1142/S0219891611002329}}, \href
  {http://dx.doi.org/10.1142/S0219891611002329}
  {\path{doi:10.1142/S0219891611002329}}.
\newline\urlprefix\url{https://doi.org/10.1142/S0219891611002329}

\bibitem{bouchut:2002}
F.~Bouchut, Entropy satisfying flux vector splittings and kinetic {BGK} models,
  Numerische Mathematik 94~(4) (2002) 623--672.

\bibitem{klingenberg:2010}
C.~Klingenberg, K.~Waagan, Relaxation solvers for ideal mhd equations -a
  review, Acta Mathematica Scientia 30 (2010) 621--632.

\bibitem{van_leer}
B.~van Leer, On the relation between the upwind-differencing schemes of
  godunov, engquist–osher and roe, SIAM Journal on Scientific and Statistical
  Computing 5 (1984) 1--20.
\newblock \href {http://dx.doi.org/10.1137/0905001}
  {\path{doi:10.1137/0905001}}.

\bibitem{DW}
W.~Dai, P.~Woodward, An approximate riemann solver for ideal
  magnetohydrodynamics, Journal of Computational Physics 111~(2) (1994)
  354--372.
\newblock \href {http://dx.doi.org/10.1006/jcph.1994.1069}
  {\path{doi:10.1006/jcph.1994.1069}}.

\bibitem{BRIO1988400}
M.~Brio, C.~Wu,
  \href{https://www.sciencedirect.com/science/article/pii/0021999188901209}{An
  upwind differencing scheme for the equations of ideal magnetohydrodynamics},
  Journal of Computational Physics 75~(2) (1988) 400--422.
\newblock \href
  {http://dx.doi.org/https://doi.org/10.1016/0021-9991(88)90120-9}
  {\path{doi:https://doi.org/10.1016/0021-9991(88)90120-9}}.
\newline\urlprefix\url{https://www.sciencedirect.com/science/article/pii/0021999188901209}

\bibitem{Falle}
S.~A. E.~G. Falle, S.~S. Komissarov, P.~Joarder,
  \href{https://doi.org/10.1046/j.1365-8711.1998.01506.x}{{A multidimensional
  upwind scheme for magnetohydrodynamics}}, Monthly Notices of the Royal
  Astronomical Society 297~(1) (1998) 265--277.
\newblock \href
  {http://arxiv.org/abs/https://academic.oup.com/mnras/article-pdf/297/1/265/18408420/297-1-265.pdf}
  {\path{arXiv:https://academic.oup.com/mnras/article-pdf/297/1/265/18408420/297-1-265.pdf}},
  \href {http://dx.doi.org/10.1046/j.1365-8711.1998.01506.x}
  {\path{doi:10.1046/j.1365-8711.1998.01506.x}}.
\newline\urlprefix\url{https://doi.org/10.1046/j.1365-8711.1998.01506.x}

\bibitem{Miyoshi}
T.~Miyoshi, K.~Kusano, A multi-state hll approximate riemann solver for ideal
  magnetohydrodynamics, Journal of Computational Physics 208 (2005) 315--344.
\newblock \href {http://dx.doi.org/10.1016/j.jcp.2005.02.017}
  {\path{doi:10.1016/j.jcp.2005.02.017}}.

\bibitem{orszag_tang_1979}
S.~A. Orszag, C.-M. Tang, Small-scale structure of two-dimensional
  magnetohydrodynamic turbulence, Journal of Fluid Mechanics 90~(1) (1979)
  129–143.
\newblock \href {http://dx.doi.org/10.1017/S002211207900210X}
  {\path{doi:10.1017/S002211207900210X}}.

\bibitem{STONE2009139}
J.~M. Stone, T.~Gardiner,
  \href{https://www.sciencedirect.com/science/article/pii/S1384107608000754}{A
  simple unsplit godunov method for multidimensional mhd}, New Astronomy 14~(2)
  (2009) 139--148.
\newblock \href
  {http://dx.doi.org/https://doi.org/10.1016/j.newast.2008.06.003}
  {\path{doi:https://doi.org/10.1016/j.newast.2008.06.003}}.
\newline\urlprefix\url{https://www.sciencedirect.com/science/article/pii/S1384107608000754}

\bibitem{BALSARA20127504}
D.~S. Balsara,
  \href{https://www.sciencedirect.com/science/article/pii/S0021999112000629}{Self-adjusting,
  positivity preserving high order schemes for hydrodynamics and
  magnetohydrodynamics}, Journal of Computational Physics 231~(22) (2012)
  7504--7517.
\newblock \href {http://dx.doi.org/https://doi.org/10.1016/j.jcp.2012.01.032}
  {\path{doi:https://doi.org/10.1016/j.jcp.2012.01.032}}.
\newline\urlprefix\url{https://www.sciencedirect.com/science/article/pii/S0021999112000629}

\bibitem{vides:2013}
{Vides, J.}, {Audit, E.}, {Guillard, H.}, {Nkonga, B.},
  \href{https://doi.org/10.1051/proc/201343012}{Divergence-free mhd simulations
  with the heracles code}, ESAIM: Proc. 43 (2013) 180--194.
\newblock \href {http://dx.doi.org/10.1051/proc/201343012}
  {\path{doi:10.1051/proc/201343012}}.
\newline\urlprefix\url{https://doi.org/10.1051/proc/201343012}

\bibitem{gonzalez:2007}
M.~{Gonz{\'a}lez}, E.~{Audit}, P.~{Huynh}, {HERACLES: a three-dimensional
  radiation hydrodynamics code}, A\&A 464~(2) (2007) 429--435.
\newblock \href {http://dx.doi.org/10.1051/0004-6361:20065486}
  {\path{doi:10.1051/0004-6361:20065486}}.

\bibitem{BALSARA1999270}
D.~S. Balsara, D.~S. Spicer,
  \href{https://www.sciencedirect.com/science/article/pii/S0021999198961538}{A
  staggered mesh algorithm using high order godunov fluxes to ensure solenoidal
  magnetic fields in magnetohydrodynamic simulations}, Journal of Computational
  Physics 149~(2) (1999) 270--292.
\newblock \href {http://dx.doi.org/https://doi.org/10.1006/jcph.1998.6153}
  {\path{doi:https://doi.org/10.1006/jcph.1998.6153}}.
\newline\urlprefix\url{https://www.sciencedirect.com/science/article/pii/S0021999198961538}

\bibitem{toth1996comparison}
G.~T{\'o}th, D.~Odstr{\v{c}}il, Comparison of some flux corrected transport and
  total variation diminishing numerical schemes for hydrodynamic and
  magnetohydrodynamic problems, Journal of Computational Physics 128~(1) (1996)
  82--100.

\end{thebibliography}

\end{document}